\documentclass[aps,prl,twocolumn,superscriptaddress]{revtex4-2}

\usepackage{graphicx}%
\usepackage{multirow}%
\usepackage{amsmath,amssymb,amsfonts}%
\usepackage{bm}%
\usepackage{amsthm,ulem}%
\usepackage{mathrsfs}%
\usepackage[title]{appendix}%
\usepackage{xcolor}%
\usepackage{textcomp}%
\usepackage{booktabs}%
\usepackage{algorithm}%
\usepackage{algorithmicx}%
\usepackage{algpseudocode}%
\usepackage{listings}%
\usepackage[separate-uncertainty=true, multi-part-units=single]{siunitx}
\usepackage{import}
\usepackage{hyperref}
\hypersetup{colorlinks,breaklinks,
	urlcolor=blue,
	linkcolor=blue,
	citecolor=blue}

\begin{document}


\title{Emulating microbial run-and-tumble and tactic motion by stochastically reorienting synthetic active Brownian particles}

\author{Sandip Kundu}
\affiliation{Department of Physics, Indian Institute of Technology Kanpur, Kanpur - 208016, India}

\author{Dibyendu Mondal}
\affiliation{Department of Physics, Indian Institute of Technology Kanpur, Kanpur - 208016, India}

\author{Arup Biswas }
\affiliation{The Institute of Mathematical Sciences, CIT Campus, Taramani, Chennai - 600113, India}
\affiliation{Homi Bhabha National Institute, Training School Complex, Anushakti Nagar, Mumbai 400094, India}

\author{Arnab Pal}
\email[Contact author: ]{arnabpal@imsc.res.in}
\affiliation{The Institute of Mathematical Sciences, CIT Campus, Taramani, Chennai - 600113, India}
\affiliation{Homi Bhabha National Institute, Training School Complex, Anushakti Nagar, Mumbai 400094, India}

\author{Manas Khan}
\email[Contact author: ]{mkhan@iitk.ac.in}
\affiliation{Department of Physics, Indian Institute of Technology Kanpur, Kanpur - 208016, India}


\begin{abstract}
Replicating efficient and adaptable microbial navigation strategies, such as run and tumble (RnT) and tactic motions to synthetic active agents has been an enduring quest. To this end, we introduce a stochastic orientational reset (SOR) protocol, in which the propulsion direction of an active Brownian particle (ABP) is reassigned to a random orientation within a defined \textit{reset-cone}. When the \textit{reset-cone} is aligned with the instantaneous propulsion direction, ABPs reproduce the RnT dynamics of \textit{E. coli}; when set along an attractant gradient, they exhibit taxis - with extensive adaptability in persistence through the angular width of the \textit{reset-cone} and reset rate. We establish the robustness of this protocol across a broad range of swimming speeds using experiments, simulations, and analytical theory. 
\end{abstract}

\maketitle

\paragraph{Introduction.}
Understanding the nonequilibrium self-propelled dynamics of microorganisms and emulating them with synthetic active particles for diverse applications has garnered significant interest over the past two decades \cite{BECKER2003, Dreyfus2005, Sanchez2014, Rao2015, Elgeti2015a, Bechinger2016, Ahmed2017, Huang2019, Liu2021, Xia2024, Anchutkin2024, Pramanik2025}. Arguably, the simplest yet most efficient natural navigation mechanism is the run-and-tumble (RnT) motion of microbes, such as \textit{E. coli} bacteria and \textit{Chlamydomonas} algae. They follow a nearly straight path at a constant speed in a “run” phase, interrupted by sudden change in the direction between consecutive runs, called “tumbles” \cite{HOWARDC.1972, Seyrich2018, Berg2004}. Wide variations in RnT motion have been observed across species and strains to adapt to their environments, such as the run-and-reverse motion of marine bacteria \cite{Johansen2002,Xie2011,Alirezaeizanjani2020,Guseva2025}. In contrast, synthetic microswimmers, most commonly realized using phoretically active half-metal-coated Janus colloids \cite{Howse2007,Jiang2010,Volpe2011,Halder2025}, move at an almost constant speed along their intrinsic direction of self-propulsion that evolves smoothly following orientational diffusion, in addition to translational Brownian fluctuations, and are called active Brownian particles (ABPs) \cite{Romanczuk2012,Bechinger2016, Fodor2018,Basu2018}. Both the RnT and ABP dynamics are ballistic at short times and eventually become diffusive-like at times longer than the characteristic timescales over which the directional correlation decays \cite{Cates2013,Solon2015,Bechinger2016}.

Despite the resemblance in long-time diffusive dynamics, the RnT and ABP motions are characteristically and significantly different in their realizations, which becomes apparent in many emergent phenomena \cite{Cates2013,Solon2015,Khatami2016}. Tumbles, \textit{i.e.}, abrupt reorientations after exponentially distributed run durations $t_{\mathrm{run}}$, as opposed to a much slower and continuous change in the direction of ABPs, provide a strategic advantage in the exploration of space with RnT \cite{Matthaeus2009, Zaburdaev2015, Rupprecht2016, Kirkegaard2018}. The average tumbling rate $\alpha$ regulates the sparsity of exploration, quantitatively defined by the long-time effective diffusion coefficient $D_{\mathrm{eff}}$, which varies as $1/\alpha$, whereas the $D_{\mathrm{eff}}$ of an ABP is set by its orientational diffusion coefficient, and hence, its size, at a given propulsion speed \cite{Nash2010,Cates2013, Khatami2016,Santra2020,mallikarjun2023chiral,Zhang2022,Romanczuk2012,Fodor2018,Basu2018}. Recent studies have revealed that the chemotaxis pathway noise-induced RnT dynamics of \textit{E. coli} resemble the L\'evy walk, which is the optimal search strategy for widely dispersed resources \cite{Matthaeus2009,Zaburdaev2015,Huo2021}. Furthermore, RnT dynamics can also manifest a biased random walk towards a more favorable environment, called tactic motion, such as chemotaxis by \textit{E. coli} and phototaxis by \textit{Chlamydomonas} \cite{HOWARDC.1972,Matthaeus2009, Saragosti2011,Taktikos2013, Kirkegaard2018,Colin2021, Raza2023,Wang2025}. Therefore, the RnT navigation strategy is considerably more efficient and flexible.

The propulsion direction of almost all artificial self-propelled particles evolves gradually because of orientational diffusion. Hence, the persistence of their dynamics is non-tunable and comparatively longer than the RnT motion, where tumbles, in addition to spontaneous orientational diffusion during runs, shorten the persistence time. Emergent RnT-resembling dynamics have been observed in a few synthetic active systems owing to asymmetric friction or intermittent constraints \cite{Lozano2018,Anchutkin2024,Paramanick2025}. However, realizing RnT dynamics with desired dynamical properties and versatility using synthetic microswimmers by applying tumble-like reorientations at a tunable rate and with an adaptable bias remains an important frontier for applications and fundamental understanding.

\begin{figure*}[htbp]
	\includegraphics[width=1\textwidth]{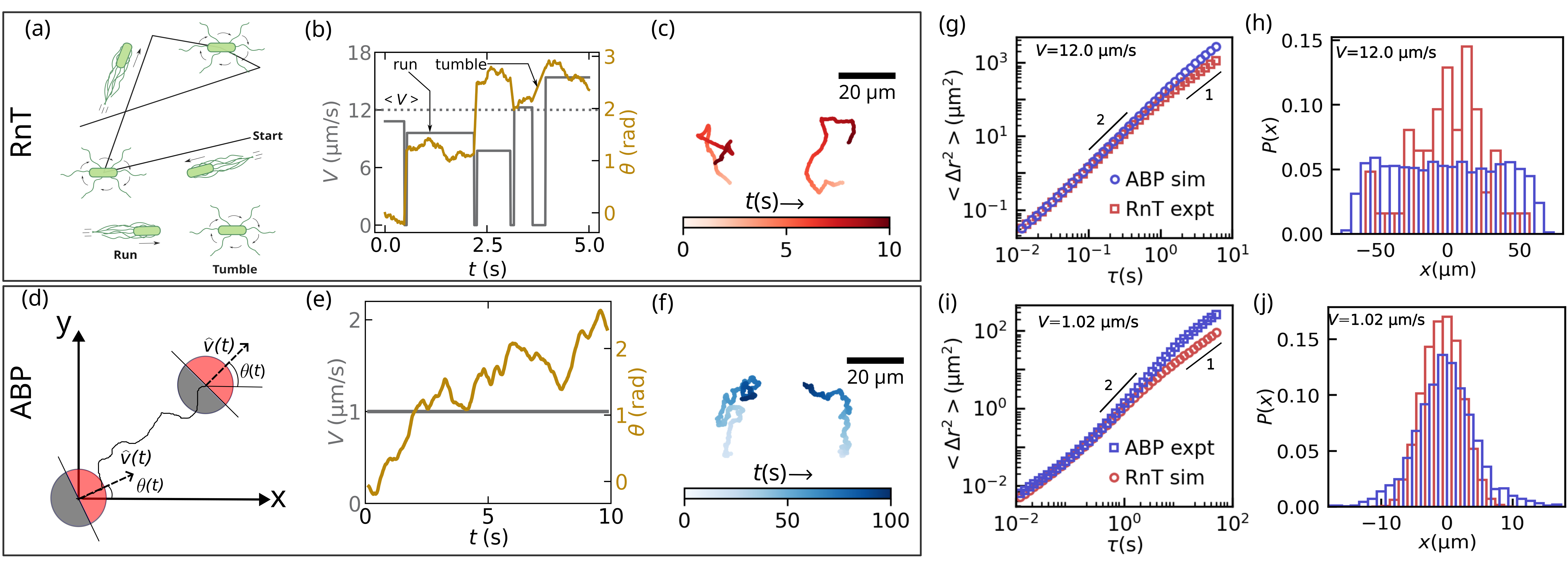}
	\caption{Characteristic differences between the RnT motion of \textit{E. coli} (a-c) and the ABP dynamics of a diffusiophoretically active Janus colloid (d-f). (a) Schematics (not to scale) show flagellar configurations of \textit{E. coli} leading to sharp reorientations between runs, called tumbles, in contrast to (d) the smoothly evolving direction of propulsion of an ABP. (b, e) Characteristic differences in their active dynamics are shown with typical variations in their speeds ($V (t)$) and directions ($\theta (t)$), and (c, f) experimentally captured typical trajectories. (g - j) Corresponding MSDs, $\left\langle \Delta r^2 \right\rangle $, and position distributions, $P(x)$, are compared for two different average propulsion speeds. (g, h) $\left\langle \Delta r^2 \right\rangle $ and $P(x)$ at \SI{6}{\s} from experimentally captured \textit{E. coli} trajectories with $\left\langle V\right\rangle $ = \SI{12.0}{\um/\s} is shown with those from simulated ABP dynamics considering the same propulsion speed. (i, j) Similarly, $\left\langle \Delta r^2 \right\rangle $ and $P(x)$ at \SI{50}{\s} from experimentally captured trajectories of an active colloid with $V$ = \SI{1.02}{\um/\s} are compared with those from simulated RnT dynamics with the same average swimming speed. (h, j) Red and blue bars denote RnT and ABP dynamics, respectively.
	\label{fig:ABPvsRnT}}
\end{figure*}

In this Letter, we report a novel generic stochastic orientational reset (SOR) protocol that incorporates tumble-like reorientations into ABP dynamics to emulate RnT motions, including its biased adaptation, \textit{i.e.}, taxis. In stark contrast to standard resetting protocols \cite{Kumar2020,sar2023resetting,Baouche2024,Paramanick2024a,Baouche2025,Shee2025}, where the orientation is always reset to a fixed direction, we reorient the propulsion direction of an ABP to a random direction drawn from a uniform distribution bounded by a cone of angular width $2\phi$ (henceforth termed as \textit{reset-cone}), which is dynamically oriented along its instantaneous orientation, intermittently after random time intervals, $\Delta t_{\mathrm{reset}}$. Notably, while directing the \textit{reset-cone} along the instantaneous propulsion direction of the ABP prior to each SOR imitates the unbiased RnT dynamics of \textit{E. coli}, anisotropic taxis motion is replicated by keeping the direction of the \textit{reset-cone} fixed along the gradient of the attractant. Our experiments, combined with extensive theory and simulations posit broad adaptability of this SOR protocol to various optimized RnT swimming strategies and tactic motion of microorganisms through the variation of $\phi$ and the reset rate $\lambda = 1/\left\langle \Delta t_{\mathrm{reset}}\right\rangle $.

\begin{figure*}[htbp]
	\includegraphics[width=\textwidth]{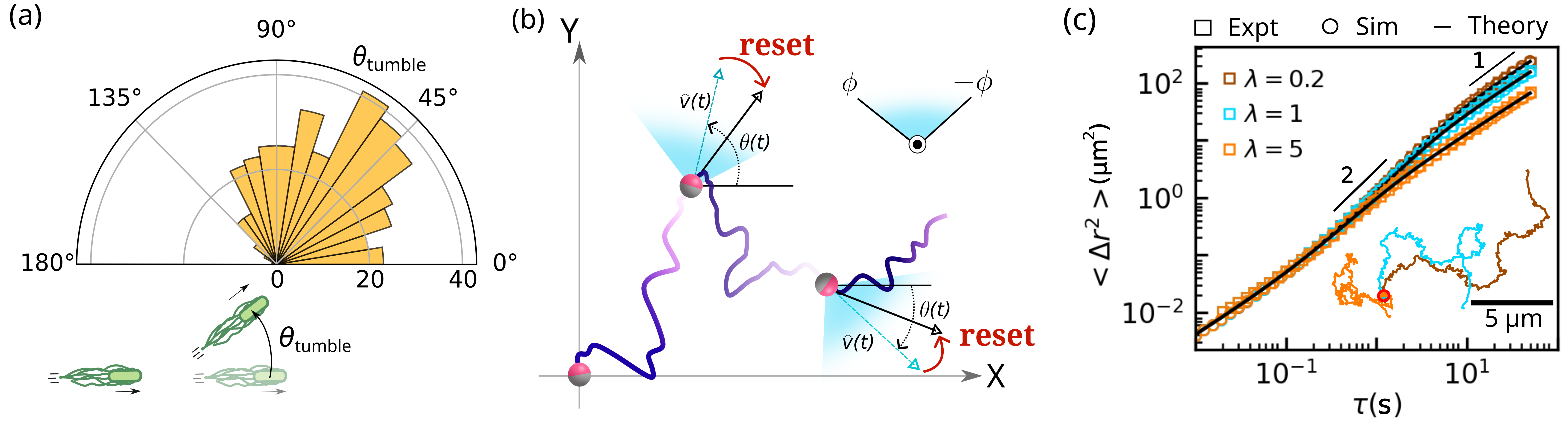}
	\caption{Stochastic orientational reset (SOR) protocol for emulating RnT dynamics. (a) Half-rose diagram showing the distribution of \textit{E. coli} tumble angles $\theta_{\mathrm{tumble}}$. (b) A typical trajectory of an active Janus colloid (grey-red sphere) is shown in a schematic with segments of purple gradients, which become lighter with time, approaching the SOR when the propulsion direction ($\hat{V} (t)$,  cyan dashed arrow) is reoriented by an angle $\theta_{\mathrm{reset}}$ (red arrow) randomly chosen from a uniform distribution bounded by a \textit{reset-cone} (cyan gradient) of angular width $2 \phi$ (inset) around $\hat{V} (t)$. (c) Three experimentally captured ABP trajectories (starting at the red circle) treated by this SOR protocol and the corresponding MSDs (squares) are shown for varying reset rates $\lambda$ with different colors. The respective MSDs from the simulated ABP dynamics (circles of the same colors) and analytical predictions (Eq. \ref{eq:MSD-th}, black lines) are superimposed.
		\label{fig:OR}}
\end{figure*}

\paragraph{RnT dynamics of E. coli.} 
As a reference for RnT dynamics, we considered the most extensively studied motion of wild-type \textit{E. coli} (strain RP437) \cite{HOWARDC.1972,Berg2004}. A fresh culture of \textit{E. coli} (SM) was studied in motility buffer at dilute concentrations, where very few bacteria were present in the field of view. The recorded RnT trajectories of the bacteria were analyzed to identify the run and tumble phases and obtain detailed statistical properties of the dynamics (Fig. \ref{fig:ABPvsRnT}(a-c, g, h)) (EM). Notably, tumbles are not instantaneous reorientations, as considered in many theoretical studies, rather take a finite time $t_{\mathrm{tumble}}$ \cite{HOWARDC.1972,Berg2004,Seyrich2018,Lee2019}. The mean swimming speed was obtained as $\left\langle V \right\rangle $ = \SI{12.0}{\um/\s}, $t_{\mathrm{run}}$ and $t_{\mathrm{tumble}}$ followed exponential distributions with mean values of \SI{0.88}{\s} and \SI{0.11}{\s}, respectively; the average tumble angle $\left\langle  \theta_{\mathrm{tumble}} \right\rangle \approx$ \ang{65} had a bias toward smaller angles, \textit{i.e.}, the forward swimming direction (Fig. \ref{fig:OR}(a)). These observations are in excellent agreement with those reported previously  \cite{HOWARDC.1972,Seyrich2018,Berg2004}.

To check the validity of our premise for widely varying swimming speeds, we further simulated the RnT dynamics at a lower $\left\langle V \right\rangle $ = \SI{1.02}{\um/\s}, resembling the propulsion speed of ABP, and with the experimentally observed distributions of $t_{\mathrm{run}}$, $t_{\mathrm{tumble}}$, and $ \theta_{\mathrm{tumble}}$ (EM). The MSD and position distribution computed from the simulated trajectories are shown in Fig. \ref{fig:ABPvsRnT}(i) and (j), respectively.

\paragraph{ABP dynamics.} 
Active Brownian dynamics were experimentally realized using half-platinum-coated silica (Pt-silica) Janus colloids with a diameter of \SI{1.76}{\um} in an aqueous solution of 4\% ($v/v$) $\mathrm{H}_2\mathrm{O}_2$ (EM, Fig. S2) \cite{Howse2007, Halder2025}. The trajectories of the diffusiophoretically active Janus particles were recorded and analyzed to obtain the statistical properties of the ABP dynamics (Fig. \ref{fig:ABPvsRnT}(d-f, i, j)), where the averaged MSD (Fig. \ref{fig:ABPvsRnT}(i)) provides the propulsion speed $V$ = \SI{1.02}{\um/\s} and persistence time $\tau_{\mathrm{R}}$ = \SI{2.61}{\s} through fitting with the analytical prediction (Eq. \ref{eq:ABP-MSD}, EM).

For a reasonable comparison with and emulation of the RnT motion of \textit{E. coli}, ABP trajectories were simulated at a higher propulsion speed of $V$ = \SI{12.0}{\um/\s} using the Langevin equations
\begin{equation}
		\frac{d \bm{r}}{dt} = V \bm{u}(\theta) + \sqrt{2D_{\mathrm{T}}} \bm{\xi}(t), \ \text{and} \ \frac{d\theta}{dt} = \sqrt{2D_{\mathrm{R}}} \eta(t),
			\label{eq:LE}
	\end{equation}
where $\bm{u}(\theta)= (cos{\theta}, sin{\theta})$ is the intrinsic direction of propulsion of the ABP, $\bm{\xi}(t) = (\xi_{x}(t),\xi_{y}(t))$ and $\eta(t)$ are independent Gaussian white noise, and $D_{\mathrm{T}}$ and $D_{\mathrm{R}}$ are the translational and orientational diffusion coefficients, respectively. $\left\langle \Delta r^2 \right\rangle $ and $P(x)$ computed from the simulated trajectories are shown in Fig. \ref{fig:ABPvsRnT}(g, h).

\paragraph{Characteristic differences between ABP and RnT dynamics.}
The exponents of the MSDs change from 2 to 1 at a longer time-lag ($\tau$) for the ABP compared to the RnT dynamics at both $V$ values (Fig. \ref{fig:ABPvsRnT}(g, i)), indicating that the ABP has longer persistence. This is further corroborated by the longer tails of the corresponding positional distributions of the ABP (Fig. \ref{fig:ABPvsRnT}(h, j)). The shorter persistence, and hence the reduced long-term effective diffusivity of the RnT, is attributed to the frequent abrupt reorientations through tumbles, which provides the advantage of optimizing the thoroughness of exploration against long-time speed.

\paragraph{SOR protocol.}
To emulate this unique dynamical property of the RnT, we apply a novel SOR to the ABP trajectories by reorienting the propulsion to a new direction with an angular displacement $\theta_{\mathrm{reset}}$, which is chosen randomly from a uniform distribution bounded by a \textit{reset-cone} of angular width $2\phi$, after exponentially distributed time intervals $\Delta t_{\mathrm{reset}}$, with reset-rate $\lambda$ (EM). Inspired by the forward bias of $\theta_{\mathrm{tumble}}$ (Fig \ref{fig:OR}(a)), we directed the \textit{reset-cone} along the instantaneous direction of the ABP prior to each SOR (Fig. \ref{fig:OR}(b)). Thereby, our SOR protocol reduces the persistence of the ABP by a required extent to match that of any pertinent RnT dynamics with suitable values of $\phi$ and $\lambda$ (EM).

\paragraph{Theoretical results.}
To account for these SOR events theoretically, we examine the conditional probability density function $P_\lambda(\theta,t|\theta_0,t_0)$ of the orientation $\theta(t)$ at time $t$ given its initial orientation $\theta_0$ at time $t_0$ and write a renewal evolution equation
\begin{align}
P_\lambda(\theta,t|\theta_0,t_0)&=e^{-\lambda (t-t_0)} P(\theta,t|\theta_0,t_0) \nonumber \\
&+ \lambda\int_{t_0}^t dt_{\mathrm{R1}}  e^{-\lambda t_{\mathrm{R1}}} P_\lambda(\theta,t|\Theta,t_{\mathrm{R1}}), \label{ren-2}
\end{align}
where the first term on the RHS accounts for the events that did not experience any resetting events during the entire duration. This is given by the probability of no resetting event upto time $t$ given by $e^{-\lambda (t-t_0)}$ multiplied by $P(\theta,t|\theta_0,t_0)$ -- the probability density of the orientation angle conditioned on no resetting event. The second term, on the other hand, computes the contribution for multiple resetting events. There, we assume that a first resetting has occurred at time $t_{\mathrm{R1}}$ so that $\theta({t_{\mathrm{R1}}}) \to \theta + \theta_{\mathrm{reset}} \equiv \Theta$, which is a random orientation angle that is uniformly chosen from the \textit{reset-cone}, specified by the interval $[\theta({t_{\mathrm{R1}}}) - \phi, \theta({t_{\mathrm{R1}}}) + \phi]$ around the instantaneous orientation angle $\theta({t_{\mathrm{R1}}})$, following which the angular dynamics renews (SM S2). Eq. (\ref{ren-2}) can be solved exactly in terms of the displacement angle $\theta(t)-\theta_0$ using standard renewal formalism \cite{evans_diffusion_2011,evans_stochastic_2020,pal2024random,tal2020experimental,besga2020optimal,Kumar2020,Baouche2025} in Fourier-Laplace space for any $\phi$. Skipping details from SM, the MSD, which is of our key interest, is obtained as
\begin{align}
	\langle \Delta r^2(\tau) \rangle= 4D_{\mathrm{T}} \tau + 2 V ^2 \tau_\lambda^2 \left( \tau / \tau_\lambda + e^{- \tau/ \tau_\lambda} -1 \right),
	\label{eq:MSD-th}
\end{align}
where $\tau_\lambda = \left( D_{\mathrm{R}}+\lambda-\lambda \frac{\sin(\phi)}{\phi} \right)^{-1}$ is SOR dependent emergent persistence time. Therefore, the relative reduction in the persistence rendered by the SOR is given by
\begin{equation}
	\Delta \tau_{\mathrm{R}} / \tau_{\mathrm{R}} = (\tau_{\mathrm{R}} - \tau_\lambda)/ \tau_{\mathrm{R}} = 1 - \tau_\lambda / \tau_{\mathrm{R}}, \label{eq: RelPersistence}
\end{equation}
whose variation with $\lambda$ for various $\phi$ is shown in Fig. \ref{fig:VariationPhiLambda}(a) (EM).

\begin{figure}[!t]
	\includegraphics[width=0.5\textwidth]{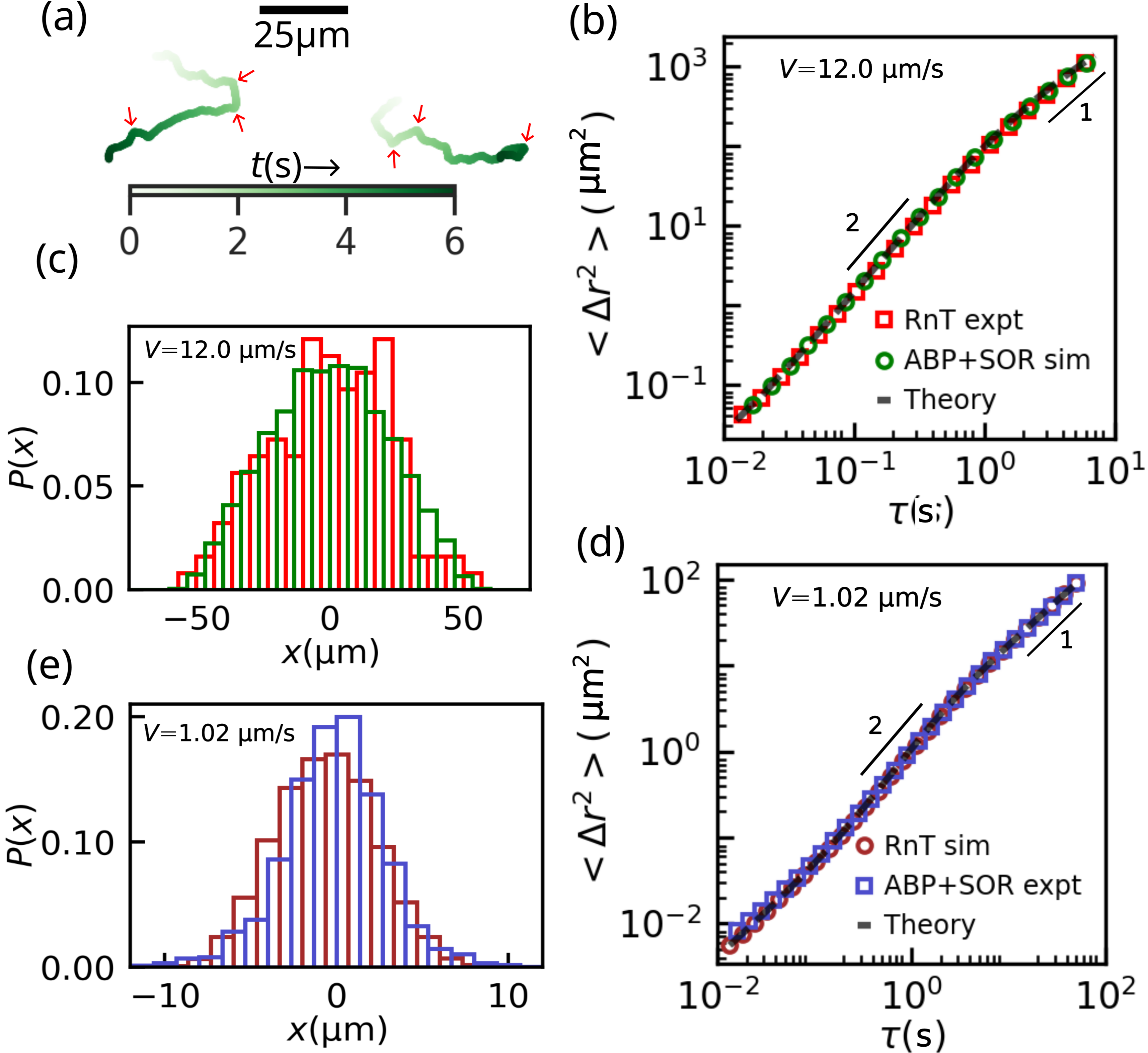}
	\caption{Emulating RnT dynamics of \textit{E. coli} by ABP motion under SOR with $\phi$ = \ang{90}, for swimming speed $V$ = \SI{12.0}{\um/\s} at $\lambda$ = 2.5 (a -c), and $V$ = \SI{1.02}{\um/\s} at $\lambda$ = 3 (d, e).  (a) Two typical resultant trajectories are shown, where a few tumble-like features are indicated by red arrows. The corresponding (b) MSD (green circles) and (c) position distribution after \SI{6}{\s} (green bars) are compared with those obtained from experimentally captured \textit{E. coli} dynamics (red squares and bars, respectively). (d) MSD (blue squares) and (e) position distribution after \SI{10}{\s} (blue bars) of the resultant motion are shown alongside those obtained from simulated RnT dynamics (brown circles and bars, respectively). (b, d) Theoretical predictions of the MSDs (Eq. \ref{eq:MSD-th}, black dashed lines) are superimposed. 
		\label{fig:RnT}}
\end{figure}

\paragraph{Shortening the persistence of ABP dynamics.}
The trajectories and corresponding MSDs obtained from experimentally captured ABP motion ($V$ = \SI{1.02}{\um/\s}, $\tau_{\mathrm{R}}$ = \SI{2.61}{\s}) under SOR with $\phi = \pi/2$ and varied $\lambda$ exhibit excellent agreement with those obtained from ABP simulations with the same SOR parameters and analytical predictions (Eq. \ref{eq:MSD-th}), demonstrating that an increasing reset rate makes the resultant dynamics less persistent, as expressed by Eq. \ref{eq: RelPersistence} (Fig. \ref{fig:OR}(c), \ref{fig:VariationPhiLambda}(a)). This resembles the variation of RnT MSD with the tumble-rate $\alpha$ \cite{Nash2010,Cates2013,Santra2020,Zhang2022}.

\paragraph{Emulating RnT dynamics.}
For a generic value of $\phi = \pi/2$, our SOR protocol emulates the RnT dynamics of \textit{E. coli} with $\lambda$ = 2.5 for a higher propulsion speed $V$ = \SI{12.0}{\um/\s}, and with $\lambda$ = 3 for a lower value of $V$ = \SI{1.02}{\um/\s}. The typical resultant trajectories resemble RnT dynamics with tumble-like events (Fig. \ref{fig:RnT}(a)). MSDs (Fig. \ref{fig:RnT}(b, d)) and position distributions (Fig. \ref{fig:RnT}(c, e)) match perfectly with those of the RnT at both swimming speeds. Furthermore, the theoretical prediction of MSD (Eq. \ref{eq:MSD-th}), indicating a reduction in persistence as given by Eq. \ref{eq: RelPersistence}, also shows excellent agreement (Fig. \ref{fig:RnT}(b, d)). Intriguingly, our SOR protocol can also replicate RnT dynamics with $\phi = \pi$, \textit{i.e.}, without any bound or bias on the reset angle. However, this enhanced randomness induces a stronger reduction in the persistence of ABP dynamics, consistent with Eq. \ref{eq: RelPersistence}, as $\phi = \pi$ reduces $\sin(\phi)/\phi$ to zero, and hence requires a smaller value of $\lambda$ (= 1) for both $V$ values (EM, Fig. \ref{fig:VariationPhiLambda}(d)).

\begin{figure}[htb]
	\includegraphics[width=0.5\textwidth]{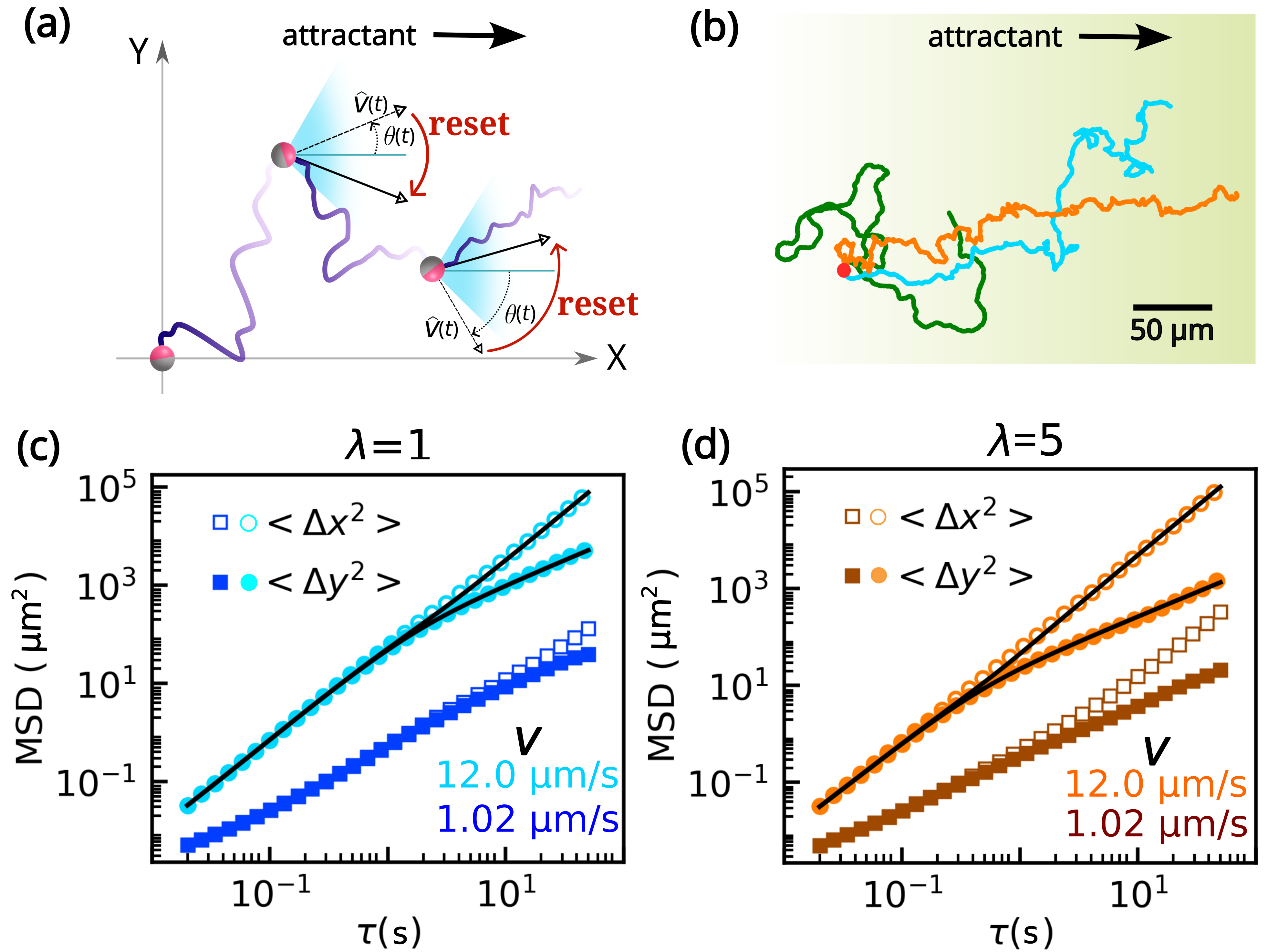}
	\caption{Imitating tactic dynamics by aligning reset cone along attractant gradient, $\hat{x}$. (a) A schematic similar to that in Fig. \ref{fig:OR}(b) shows the SOR protocol for emulating taxis. (b) Two resultant trajectories of \SI{50}{\s} duration ($V$ = \SI{12.0}{\um/\s}) under this SOR protocol with $\lambda$ = 1 (cyan) and 5 (orange), are exhibited alongside the one with the RnT-mimicking SOR protocol (green), starting at the red dot. 1D MSDs along $\hat{x}$ (open symbols) and $\hat{y}$ (filled symbols) are shown at two different reset rates: (c) $\lambda$ = 1 and (d) $\lambda$ = 5, for both $V$ = \SI{12.0}{\um/\s} (cyan and orange, respectively) and \SI{1.02}{\um/\s} (blue and brown, respectively). (c, d) The theoretical predictions (black lines) are superimposed on the MSDs corresponding to $V$ = \SI{12.0}{\um/\s}. 
		\label{fig:Chemotaxis}}
\end{figure}

\paragraph{Emulating tactic dynamics.}
Further emphasizing the generality of our SOR protocol, we show that it can emulate the biased tactic dynamics of microorganisms responding to the gradient of an attractant simply by orienting the \textit{reset-cone} along the direction of bias, considered here to be $\hat{x}$ (Fig. \ref{fig:Chemotaxis}(a)). To demonstrate this, we use the same generic value of $\phi = \pi/2$ and show that the net directional speed of the resultant dynamics, \textit{i.e.}, the tactic efficiency, increases with increasing value of $\lambda$, from 1 to 5 (Fig \ref{fig:Chemotaxis}(b)). Long-time directionality in the resultant motion is also evident in the divergence between the 1D MSDs along the gradient of the attractant ($\left\langle \Delta x^2 \right\rangle$) and perpendicular to it ($\left\langle \Delta y^2 \right\rangle$), which increases with increasing $\lambda$ and $V$. Both MSDs corresponding to $V$ = \SI{12.0}{\um/\s} show excellent agreement with the theoretical prediction (SM section S2D) (Fig. \ref{fig:Chemotaxis}(c, d)). Therefore, tactic efficiency is enhanced by more frequent reorientations with a bias along the attractant gradient.

\paragraph{Conclusions.}
In this \textit{Letter}, we have established that the dynamics of active Brownian particles (ABPs) governed by the generic SOR protocols can emulate run-and-tumble (RnT) motion, as well as its biased adaptation, \textit{i.e.}, taxis. The versatility of this approach is further demonstrated by its successful validation across two distinct and widely separated mobility regimes. The orientation of the reset-cone governs the long-term persistence, leading to an isotropic RnT motion with stochastically varying direction attached to the instantaneous propulsion $\hat{V}(t)$ and a directed tactic motion with a fixed orientation along the attractant gradient $\hat{x}$, while its angular width $2\phi$ and reset rate $\lambda$ regulate the shortening of the persistence of the resultant dynamics. An increase in $\phi$ enhances the randomness in the reset angle and hence decreases the persistence more efficiently, requiring less frequent resets, whereas a higher value of $\lambda$ shortens the persistence with finer control to match that of the desired RnT variation (EM, Fig. \ref{fig:VariationPhiLambda}).

In addition to strengthening our understanding of how SOR alters the persistence of active dynamics, our findings provide a convenient way to devise artificial RnT swimmers with desired swimming speeds and navigation strategies optimized for specific environments and applications. Although we applied an orientational transformation to the subsequent part of an ABP trajectory for implementing SOR (EM), appropriately designed artificial microswimmers, such as Janus active colloids with magnetic coatings or patches \cite{Martin2013, Han2020, Bishop2023} and microrobots \cite{Ye2023, Paramanick2024}, can be directly reoriented using an external field, feedback, or programming.

Generalizing or revising our SOR protocol to emulate other microbial navigation techniques using artificial microswimmers and developing pertinent theoretical models represent important future directions. This study demonstrates that innovative resetting rules can fundamentally reshape navigation and search strategies, thereby underscoring the impetus for designing novel reset protocols.

\paragraph{Acknowledgment.}
We thank Kazumasa A. Takeuchi, The University of Tokyo, for providing the E. coli strain (RP437) used in this study, and the PARAM Sanganak computing facility at the Computer Center, IIT Kanpur, for the numerical simulations of ABP and RnT dynamics. MK acknowledges funding from the SERB (CRG/2020/002723) and MHRD (MoE-STARS/STARS-2/2023-0814) for supporting this research. DM acknowledges financial support from PMRF (PMR8017). The numerical calculations validating theoretical results were carried out on the Kamet cluster, which is maintained and supported by the Institute of Mathematical Science’s High-Performance Computing Center. AB and AP gratefully acknowledge research support from the Department of Atomic Energy, Government of India via the Apex Projects. AP acknowledges research support from the Department of Science and Technology, India, SERB Start-up Research Grant Number SRG/2022/000080. AP also acknowledges International Research Project (IRP) titled “Classical and quantum dynamics in out of equilibrium systems” by CNRS, France.






\bibliography{ABP-OR-RnT.bib}

\begin{thebibliography}{62}%
\makeatletter
\providecommand \@ifxundefined [1]{%
 \@ifx{#1\undefined}
}%
\providecommand \@ifnum [1]{%
 \ifnum #1\expandafter \@firstoftwo
 \else \expandafter \@secondoftwo
 \fi
}%
\providecommand \@ifx [1]{%
 \ifx #1\expandafter \@firstoftwo
 \else \expandafter \@secondoftwo
 \fi
}%
\providecommand \natexlab [1]{#1}%
\providecommand \enquote  [1]{``#1''}%
\providecommand \bibnamefont  [1]{#1}%
\providecommand \bibfnamefont [1]{#1}%
\providecommand \citenamefont [1]{#1}%
\providecommand \href@noop [0]{\@secondoftwo}%
\providecommand \href [0]{\begingroup \@sanitize@url \@href}%
\providecommand \@href[1]{\@@startlink{#1}\@@href}%
\providecommand \@@href[1]{\endgroup#1\@@endlink}%
\providecommand \@sanitize@url [0]{\catcode `\\12\catcode `\$12\catcode
  `\&12\catcode `\#12\catcode `\^12\catcode `\_12\catcode `\%12\relax}%
\providecommand \@@startlink[1]{}%
\providecommand \@@endlink[0]{}%
\providecommand \url  [0]{\begingroup\@sanitize@url \@url }%
\providecommand \@url [1]{\endgroup\@href {#1}{\urlprefix }}%
\providecommand \urlprefix  [0]{URL }%
\providecommand \Eprint [0]{\href }%
\providecommand \doibase [0]{https://doi.org/}%
\providecommand \selectlanguage [0]{\@gobble}%
\providecommand \bibinfo  [0]{\@secondoftwo}%
\providecommand \bibfield  [0]{\@secondoftwo}%
\providecommand \translation [1]{[#1]}%
\providecommand \BibitemOpen [0]{}%
\providecommand \bibitemStop [0]{}%
\providecommand \bibitemNoStop [0]{.\EOS\space}%
\providecommand \EOS [0]{\spacefactor3000\relax}%
\providecommand \BibitemShut  [1]{\csname bibitem#1\endcsname}%
\let\auto@bib@innerbib\@empty
\bibitem [{\citenamefont {BECKER}\ \emph {et~al.}(2003)\citenamefont {BECKER},
  \citenamefont {KOEHLER},\ and\ \citenamefont {STONE}}]{BECKER2003}%
  \BibitemOpen
  \bibfield  {author} {\bibinfo {author} {\bibfnamefont {L.~E.}\ \bibnamefont
  {BECKER}}, \bibinfo {author} {\bibfnamefont {S.~A.}\ \bibnamefont
  {KOEHLER}},\ and\ \bibinfo {author} {\bibfnamefont {H.~A.}\ \bibnamefont
  {STONE}},\ }\bibfield  {title} {\bibinfo {title} {On self-propulsion of
  micro-machines at low reynolds number: Purcell’s three-link swimmer},\
  }\href {https://doi.org/10.1017/s0022112003005184} {\bibfield  {journal}
  {\bibinfo  {journal} {Journal of Fluid Mechanics}\ }\textbf {\bibinfo
  {volume} {490}},\ \bibinfo {pages} {15} (\bibinfo {year} {2003})}\BibitemShut
  {NoStop}%
\bibitem [{\citenamefont {Dreyfus}\ \emph {et~al.}(2005)\citenamefont
  {Dreyfus}, \citenamefont {Baudry}, \citenamefont {Roper}, \citenamefont
  {Fermigier}, \citenamefont {Stone},\ and\ \citenamefont
  {Bibette}}]{Dreyfus2005}%
  \BibitemOpen
  \bibfield  {author} {\bibinfo {author} {\bibfnamefont {R.}~\bibnamefont
  {Dreyfus}}, \bibinfo {author} {\bibfnamefont {J.}~\bibnamefont {Baudry}},
  \bibinfo {author} {\bibfnamefont {M.~L.}\ \bibnamefont {Roper}}, \bibinfo
  {author} {\bibfnamefont {M.}~\bibnamefont {Fermigier}}, \bibinfo {author}
  {\bibfnamefont {H.~A.}\ \bibnamefont {Stone}},\ and\ \bibinfo {author}
  {\bibfnamefont {J.}~\bibnamefont {Bibette}},\ }\bibfield  {title} {\bibinfo
  {title} {Microscopic artificial swimmers},\ }\href
  {https://doi.org/10.1038/nature04090} {\bibfield  {journal} {\bibinfo
  {journal} {Nature}\ }\textbf {\bibinfo {volume} {437}},\ \bibinfo {pages}
  {862} (\bibinfo {year} {2005})}\BibitemShut {NoStop}%
\bibitem [{\citenamefont {Sánchez}\ \emph {et~al.}(2014)\citenamefont
  {Sánchez}, \citenamefont {Soler},\ and\ \citenamefont
  {Katuri}}]{Sanchez2014}%
  \BibitemOpen
  \bibfield  {author} {\bibinfo {author} {\bibfnamefont {S.}~\bibnamefont
  {Sánchez}}, \bibinfo {author} {\bibfnamefont {L.}~\bibnamefont {Soler}},\
  and\ \bibinfo {author} {\bibfnamefont {J.}~\bibnamefont {Katuri}},\
  }\bibfield  {title} {\bibinfo {title} {Chemically powered micro‐ and
  nanomotors},\ }\href {https://doi.org/10.1002/anie.201406096} {\bibfield
  {journal} {\bibinfo  {journal} {Angewandte Chemie International Edition}\
  }\textbf {\bibinfo {volume} {54}},\ \bibinfo {pages} {1414} (\bibinfo {year}
  {2014})}\BibitemShut {NoStop}%
\bibitem [{\citenamefont {Rao}\ \emph {et~al.}(2015)\citenamefont {Rao},
  \citenamefont {Li}, \citenamefont {Meng}, \citenamefont {Zheng},
  \citenamefont {Cai},\ and\ \citenamefont {Wang}}]{Rao2015}%
  \BibitemOpen
  \bibfield  {author} {\bibinfo {author} {\bibfnamefont {K.~J.}\ \bibnamefont
  {Rao}}, \bibinfo {author} {\bibfnamefont {F.}~\bibnamefont {Li}}, \bibinfo
  {author} {\bibfnamefont {L.}~\bibnamefont {Meng}}, \bibinfo {author}
  {\bibfnamefont {H.}~\bibnamefont {Zheng}}, \bibinfo {author} {\bibfnamefont
  {F.}~\bibnamefont {Cai}},\ and\ \bibinfo {author} {\bibfnamefont
  {W.}~\bibnamefont {Wang}},\ }\bibfield  {title} {\bibinfo {title} {A force to
  be reckoned with: A review of synthetic microswimmers powered by
  ultrasound},\ }\href {https://doi.org/10.1002/smll.201403621} {\bibfield
  {journal} {\bibinfo  {journal} {Small}\ }\textbf {\bibinfo {volume} {11}},\
  \bibinfo {pages} {2836} (\bibinfo {year} {2015})}\BibitemShut {NoStop}%
\bibitem [{\citenamefont {Elgeti}\ \emph {et~al.}(2015)\citenamefont {Elgeti},
  \citenamefont {Winkler},\ and\ \citenamefont {Gompper}}]{Elgeti2015a}%
  \BibitemOpen
  \bibfield  {author} {\bibinfo {author} {\bibfnamefont {J.}~\bibnamefont
  {Elgeti}}, \bibinfo {author} {\bibfnamefont {R.~G.}\ \bibnamefont
  {Winkler}},\ and\ \bibinfo {author} {\bibfnamefont {G.}~\bibnamefont
  {Gompper}},\ }\bibfield  {title} {\bibinfo {title} {Physics of
  microswimmers—single particle motion and collective behavior: a review},\
  }\href {https://doi.org/10.1088/0034-4885/78/5/056601} {\bibfield  {journal}
  {\bibinfo  {journal} {Reports on Progress in Physics}\ }\textbf {\bibinfo
  {volume} {78}},\ \bibinfo {pages} {056601} (\bibinfo {year}
  {2015})}\BibitemShut {NoStop}%
\bibitem [{\citenamefont {Bechinger}\ \emph {et~al.}(2016)\citenamefont
  {Bechinger}, \citenamefont {Di~Leonardo}, \citenamefont {Löwen},
  \citenamefont {Reichhardt}, \citenamefont {Volpe},\ and\ \citenamefont
  {Volpe}}]{Bechinger2016}%
  \BibitemOpen
  \bibfield  {author} {\bibinfo {author} {\bibfnamefont {C.}~\bibnamefont
  {Bechinger}}, \bibinfo {author} {\bibfnamefont {R.}~\bibnamefont
  {Di~Leonardo}}, \bibinfo {author} {\bibfnamefont {H.}~\bibnamefont {Löwen}},
  \bibinfo {author} {\bibfnamefont {C.}~\bibnamefont {Reichhardt}}, \bibinfo
  {author} {\bibfnamefont {G.}~\bibnamefont {Volpe}},\ and\ \bibinfo {author}
  {\bibfnamefont {G.}~\bibnamefont {Volpe}},\ }\bibfield  {title} {\bibinfo
  {title} {Active particles in complex and crowded environments},\ }\href
  {https://doi.org/10.1103/revmodphys.88.045006} {\bibfield  {journal}
  {\bibinfo  {journal} {Reviews of Modern Physics}\ }\textbf {\bibinfo {volume}
  {88}},\ \bibinfo {pages} {045006} (\bibinfo {year} {2016})}\BibitemShut
  {NoStop}%
\bibitem [{\citenamefont {Ahmed}\ \emph {et~al.}(2017)\citenamefont {Ahmed},
  \citenamefont {Dillinger}, \citenamefont {Hong},\ and\ \citenamefont
  {Nelson}}]{Ahmed2017}%
  \BibitemOpen
  \bibfield  {author} {\bibinfo {author} {\bibfnamefont {D.}~\bibnamefont
  {Ahmed}}, \bibinfo {author} {\bibfnamefont {C.}~\bibnamefont {Dillinger}},
  \bibinfo {author} {\bibfnamefont {A.}~\bibnamefont {Hong}},\ and\ \bibinfo
  {author} {\bibfnamefont {B.~J.}\ \bibnamefont {Nelson}},\ }\bibfield  {title}
  {\bibinfo {title} {Artificial acousto‐magnetic soft microswimmers},\
  }\bibfield  {journal} {\bibinfo  {journal} {Advanced Materials Technologies}\
  }\textbf {\bibinfo {volume} {2}},\ \href
  {https://doi.org/10.1002/admt.201700050} {10.1002/admt.201700050} (\bibinfo
  {year} {2017})\BibitemShut {NoStop}%
\bibitem [{\citenamefont {Huang}\ \emph {et~al.}(2019)\citenamefont {Huang},
  \citenamefont {Uslu}, \citenamefont {Katsamba}, \citenamefont {Lauga},
  \citenamefont {Sakar},\ and\ \citenamefont {Nelson}}]{Huang2019}%
  \BibitemOpen
  \bibfield  {author} {\bibinfo {author} {\bibfnamefont {H.-W.}\ \bibnamefont
  {Huang}}, \bibinfo {author} {\bibfnamefont {F.~E.}\ \bibnamefont {Uslu}},
  \bibinfo {author} {\bibfnamefont {P.}~\bibnamefont {Katsamba}}, \bibinfo
  {author} {\bibfnamefont {E.}~\bibnamefont {Lauga}}, \bibinfo {author}
  {\bibfnamefont {M.~S.}\ \bibnamefont {Sakar}},\ and\ \bibinfo {author}
  {\bibfnamefont {B.~J.}\ \bibnamefont {Nelson}},\ }\bibfield  {title}
  {\bibinfo {title} {Adaptive locomotion of artificial microswimmers},\
  }\bibfield  {journal} {\bibinfo  {journal} {Science Advances}\ }\textbf
  {\bibinfo {volume} {5}},\ \href {https://doi.org/10.1126/sciadv.aau1532}
  {10.1126/sciadv.aau1532} (\bibinfo {year} {2019})\BibitemShut {NoStop}%
\bibitem [{\citenamefont {Liu}\ \emph {et~al.}(2021)\citenamefont {Liu},
  \citenamefont {Fu}, \citenamefont {Liu},\ and\ \citenamefont
  {Ruan}}]{Liu2021}%
  \BibitemOpen
  \bibfield  {author} {\bibinfo {author} {\bibfnamefont {J.}~\bibnamefont
  {Liu}}, \bibinfo {author} {\bibfnamefont {Y.}~\bibnamefont {Fu}}, \bibinfo
  {author} {\bibfnamefont {X.}~\bibnamefont {Liu}},\ and\ \bibinfo {author}
  {\bibfnamefont {H.}~\bibnamefont {Ruan}},\ }\bibfield  {title} {\bibinfo
  {title} {Theoretical perspectives on natural and artificial micro-swimmers},\
  }\href {https://doi.org/10.1007/s10338-021-00260-w} {\bibfield  {journal}
  {\bibinfo  {journal} {Acta Mechanica Solida Sinica}\ }\textbf {\bibinfo
  {volume} {34}},\ \bibinfo {pages} {783} (\bibinfo {year} {2021})}\BibitemShut
  {NoStop}%
\bibitem [{\citenamefont {Xia}\ \emph {et~al.}(2024)\citenamefont {Xia},
  \citenamefont {Hu}, \citenamefont {Wei}, \citenamefont {Chen}, \citenamefont
  {Peng},\ and\ \citenamefont {Yang}}]{Xia2024}%
  \BibitemOpen
  \bibfield  {author} {\bibinfo {author} {\bibfnamefont {Y.}~\bibnamefont
  {Xia}}, \bibinfo {author} {\bibfnamefont {Z.}~\bibnamefont {Hu}}, \bibinfo
  {author} {\bibfnamefont {D.}~\bibnamefont {Wei}}, \bibinfo {author}
  {\bibfnamefont {K.}~\bibnamefont {Chen}}, \bibinfo {author} {\bibfnamefont
  {Y.}~\bibnamefont {Peng}},\ and\ \bibinfo {author} {\bibfnamefont
  {M.}~\bibnamefont {Yang}},\ }\bibfield  {title} {\bibinfo {title} {Biomimetic
  synchronization in biciliated robots},\ }\href
  {https://doi.org/10.1103/physrevlett.133.048302} {\bibfield  {journal}
  {\bibinfo  {journal} {Physical Review Letters}\ }\textbf {\bibinfo {volume}
  {133}},\ \bibinfo {pages} {048302} (\bibinfo {year} {2024})}\BibitemShut
  {NoStop}%
\bibitem [{\citenamefont {Anchutkin}\ \emph {et~al.}(2024)\citenamefont
  {Anchutkin}, \citenamefont {Cichos},\ and\ \citenamefont
  {Holubec}}]{Anchutkin2024}%
  \BibitemOpen
  \bibfield  {author} {\bibinfo {author} {\bibfnamefont {G.}~\bibnamefont
  {Anchutkin}}, \bibinfo {author} {\bibfnamefont {F.}~\bibnamefont {Cichos}},\
  and\ \bibinfo {author} {\bibfnamefont {V.}~\bibnamefont {Holubec}},\
  }\bibfield  {title} {\bibinfo {title} {Run-and-tumble motion of ellipsoidal
  microswimmers},\ }\href {https://doi.org/10.1103/physrevresearch.6.043101}
  {\bibfield  {journal} {\bibinfo  {journal} {Physical Review Research}\
  }\textbf {\bibinfo {volume} {6}},\ \bibinfo {pages} {043101} (\bibinfo {year}
  {2024})}\BibitemShut {NoStop}%
\bibitem [{\citenamefont {Pramanik}\ \emph {et~al.}(2025)\citenamefont
  {Pramanik}, \citenamefont {Mishra},\ and\ \citenamefont
  {Chatterjee}}]{Pramanik2025}%
  \BibitemOpen
  \bibfield  {author} {\bibinfo {author} {\bibfnamefont {R.}~\bibnamefont
  {Pramanik}}, \bibinfo {author} {\bibfnamefont {S.}~\bibnamefont {Mishra}},\
  and\ \bibinfo {author} {\bibfnamefont {S.}~\bibnamefont {Chatterjee}},\
  }\bibfield  {title} {\bibinfo {title} {Run-and-tumble chemotaxis using
  reinforcement learning},\ }\href
  {https://doi.org/10.1103/physreve.111.014106} {\bibfield  {journal} {\bibinfo
   {journal} {Physical Review E}\ }\textbf {\bibinfo {volume} {111}},\ \bibinfo
  {pages} {014106} (\bibinfo {year} {2025})}\BibitemShut {NoStop}%
\bibitem [{\citenamefont {BERG}\ and\ \citenamefont
  {BROWN}(1972)}]{HOWARDC.1972}%
  \BibitemOpen
  \bibfield  {author} {\bibinfo {author} {\bibfnamefont {H.~C.}\ \bibnamefont
  {BERG}}\ and\ \bibinfo {author} {\bibfnamefont {D.~A.}\ \bibnamefont
  {BROWN}},\ }\bibfield  {title} {\bibinfo {title} {Chemotaxis in escherichia
  coli analysed by three-dimensional tracking},\ }\href
  {https://doi.org/10.1038/239500a0} {\bibfield  {journal} {\bibinfo  {journal}
  {Nature}\ }\textbf {\bibinfo {volume} {239}},\ \bibinfo {pages} {500}
  (\bibinfo {year} {1972})}\BibitemShut {NoStop}%
\bibitem [{\citenamefont {Seyrich}\ \emph {et~al.}(2018)\citenamefont
  {Seyrich}, \citenamefont {Alirezaeizanjani}, \citenamefont {Beta},\ and\
  \citenamefont {Stark}}]{Seyrich2018}%
  \BibitemOpen
  \bibfield  {author} {\bibinfo {author} {\bibfnamefont {M.}~\bibnamefont
  {Seyrich}}, \bibinfo {author} {\bibfnamefont {Z.}~\bibnamefont
  {Alirezaeizanjani}}, \bibinfo {author} {\bibfnamefont {C.}~\bibnamefont
  {Beta}},\ and\ \bibinfo {author} {\bibfnamefont {H.}~\bibnamefont {Stark}},\
  }\bibfield  {title} {\bibinfo {title} {Statistical parameter inference of
  bacterial swimming strategies},\ }\href
  {https://doi.org/10.1088/1367-2630/aae72c} {\bibfield  {journal} {\bibinfo
  {journal} {New Journal of Physics}\ }\textbf {\bibinfo {volume} {20}},\
  \bibinfo {pages} {103033} (\bibinfo {year} {2018})}\BibitemShut {NoStop}%
\bibitem [{\citenamefont {Berg}(2004)}]{Berg2004}%
  \BibitemOpen
  \bibfield  {author} {\bibinfo {author} {\bibfnamefont {H.~C.}\ \bibnamefont
  {Berg}},\ }\href {https://doi.org/https://doi.org/10.1007/b97370} {\emph
  {\bibinfo {title} {E. coli in Motion}}},\ \bibinfo {edition} {1st}\ ed.,\
  edited by\ \bibinfo {editor} {\bibfnamefont {H.~C.}\ \bibnamefont {Berg}},\
  Biological and Medical Physics, Biomedical Engineering\ (\bibinfo
  {publisher} {Springer New York},\ \bibinfo {address} {New York, NY},\
  \bibinfo {year} {2004})\BibitemShut {NoStop}%
\bibitem [{\citenamefont {Johansen}\ \emph {et~al.}(2002)\citenamefont
  {Johansen}, \citenamefont {Pinhassi}, \citenamefont {Blackburn},
  \citenamefont {Zweifel},\ and\ \citenamefont {Hagstrom}}]{Johansen2002}%
  \BibitemOpen
  \bibfield  {author} {\bibinfo {author} {\bibfnamefont {J.}~\bibnamefont
  {Johansen}}, \bibinfo {author} {\bibfnamefont {J.}~\bibnamefont {Pinhassi}},
  \bibinfo {author} {\bibfnamefont {N.}~\bibnamefont {Blackburn}}, \bibinfo
  {author} {\bibfnamefont {U.}~\bibnamefont {Zweifel}},\ and\ \bibinfo {author}
  {\bibfnamefont {A.}~\bibnamefont {Hagstrom}},\ }\bibfield  {title} {\bibinfo
  {title} {Variability in motility characteristics among marine bacteria},\
  }\href {https://doi.org/10.3354/ame028229} {\bibfield  {journal} {\bibinfo
  {journal} {Aquatic Microbial Ecology}\ }\textbf {\bibinfo {volume} {28}},\
  \bibinfo {pages} {229} (\bibinfo {year} {2002})}\BibitemShut {NoStop}%
\bibitem [{\citenamefont {Xie}\ \emph {et~al.}(2011)\citenamefont {Xie},
  \citenamefont {Altindal}, \citenamefont {Chattopadhyay},\ and\ \citenamefont
  {Wu}}]{Xie2011}%
  \BibitemOpen
  \bibfield  {author} {\bibinfo {author} {\bibfnamefont {L.}~\bibnamefont
  {Xie}}, \bibinfo {author} {\bibfnamefont {T.}~\bibnamefont {Altindal}},
  \bibinfo {author} {\bibfnamefont {S.}~\bibnamefont {Chattopadhyay}},\ and\
  \bibinfo {author} {\bibfnamefont {X.-L.}\ \bibnamefont {Wu}},\ }\bibfield
  {title} {\bibinfo {title} {Bacterial flagellum as a propeller and as a rudder
  for efficient chemotaxis},\ }\href {https://doi.org/10.1073/pnas.1011953108}
  {\bibfield  {journal} {\bibinfo  {journal} {Proceedings of the National
  Academy of Sciences}\ }\textbf {\bibinfo {volume} {108}},\ \bibinfo {pages}
  {2246} (\bibinfo {year} {2011})}\BibitemShut {NoStop}%
\bibitem [{\citenamefont {Alirezaeizanjani}\ \emph {et~al.}(2020)\citenamefont
  {Alirezaeizanjani}, \citenamefont {Großmann}, \citenamefont {Pfeifer},
  \citenamefont {Hintsche},\ and\ \citenamefont {Beta}}]{Alirezaeizanjani2020}%
  \BibitemOpen
  \bibfield  {author} {\bibinfo {author} {\bibfnamefont {Z.}~\bibnamefont
  {Alirezaeizanjani}}, \bibinfo {author} {\bibfnamefont {R.}~\bibnamefont
  {Großmann}}, \bibinfo {author} {\bibfnamefont {V.}~\bibnamefont {Pfeifer}},
  \bibinfo {author} {\bibfnamefont {M.}~\bibnamefont {Hintsche}},\ and\
  \bibinfo {author} {\bibfnamefont {C.}~\bibnamefont {Beta}},\ }\bibfield
  {title} {\bibinfo {title} {Chemotaxis strategies of bacteria with multiple
  run modes},\ }\bibfield  {journal} {\bibinfo  {journal} {Science Advances}\
  }\textbf {\bibinfo {volume} {6}},\ \href
  {https://doi.org/10.1126/sciadv.aaz6153} {10.1126/sciadv.aaz6153} (\bibinfo
  {year} {2020})\BibitemShut {NoStop}%
\bibitem [{\citenamefont {Guseva}\ and\ \citenamefont
  {Feudel}(2025)}]{Guseva2025}%
  \BibitemOpen
  \bibfield  {author} {\bibinfo {author} {\bibfnamefont {K.}~\bibnamefont
  {Guseva}}\ and\ \bibinfo {author} {\bibfnamefont {U.}~\bibnamefont
  {Feudel}},\ }\bibfield  {title} {\bibinfo {title} {Advantages of run-reverse
  motility pattern of bacteria for tracking light and small food sources in
  dynamic fluid environments},\ }\bibfield  {journal} {\bibinfo  {journal}
  {Journal of The Royal Society Interface}\ }\textbf {\bibinfo {volume} {22}},\
  \href {https://doi.org/10.1098/rsif.2025.0037} {10.1098/rsif.2025.0037}
  (\bibinfo {year} {2025})\BibitemShut {NoStop}%
\bibitem [{\citenamefont {Howse}\ \emph {et~al.}(2007)\citenamefont {Howse},
  \citenamefont {Jones}, \citenamefont {Ryan}, \citenamefont {Gough},
  \citenamefont {Vafabakhsh},\ and\ \citenamefont {Golestanian}}]{Howse2007}%
  \BibitemOpen
  \bibfield  {author} {\bibinfo {author} {\bibfnamefont {J.~R.}\ \bibnamefont
  {Howse}}, \bibinfo {author} {\bibfnamefont {R.~A.~L.}\ \bibnamefont {Jones}},
  \bibinfo {author} {\bibfnamefont {A.~J.}\ \bibnamefont {Ryan}}, \bibinfo
  {author} {\bibfnamefont {T.}~\bibnamefont {Gough}}, \bibinfo {author}
  {\bibfnamefont {R.}~\bibnamefont {Vafabakhsh}},\ and\ \bibinfo {author}
  {\bibfnamefont {R.}~\bibnamefont {Golestanian}},\ }\bibfield  {title}
  {\bibinfo {title} {Self-motile colloidal particles: From directed propulsion
  to random walk},\ }\href {https://doi.org/10.1103/physrevlett.99.048102}
  {\bibfield  {journal} {\bibinfo  {journal} {Physical Review Letters}\
  }\textbf {\bibinfo {volume} {99}},\ \bibinfo {pages} {048102} (\bibinfo
  {year} {2007})}\BibitemShut {NoStop}%
\bibitem [{\citenamefont {Jiang}\ \emph {et~al.}(2010)\citenamefont {Jiang},
  \citenamefont {Yoshinaga},\ and\ \citenamefont {Sano}}]{Jiang2010}%
  \BibitemOpen
  \bibfield  {author} {\bibinfo {author} {\bibfnamefont {H.-R.}\ \bibnamefont
  {Jiang}}, \bibinfo {author} {\bibfnamefont {N.}~\bibnamefont {Yoshinaga}},\
  and\ \bibinfo {author} {\bibfnamefont {M.}~\bibnamefont {Sano}},\ }\bibfield
  {title} {\bibinfo {title} {Active motion of a janus particle by
  self-thermophoresis in a defocused laser beam},\ }\href
  {https://doi.org/10.1103/physrevlett.105.268302} {\bibfield  {journal}
  {\bibinfo  {journal} {Physical Review Letters}\ }\textbf {\bibinfo {volume}
  {105}},\ \bibinfo {pages} {268302} (\bibinfo {year} {2010})}\BibitemShut
  {NoStop}%
\bibitem [{\citenamefont {Volpe}\ \emph {et~al.}(2011)\citenamefont {Volpe},
  \citenamefont {Buttinoni}, \citenamefont {Vogt}, \citenamefont {Kümmerer},\
  and\ \citenamefont {Bechinger}}]{Volpe2011}%
  \BibitemOpen
  \bibfield  {author} {\bibinfo {author} {\bibfnamefont {G.}~\bibnamefont
  {Volpe}}, \bibinfo {author} {\bibfnamefont {I.}~\bibnamefont {Buttinoni}},
  \bibinfo {author} {\bibfnamefont {D.}~\bibnamefont {Vogt}}, \bibinfo {author}
  {\bibfnamefont {H.-J.}\ \bibnamefont {Kümmerer}},\ and\ \bibinfo {author}
  {\bibfnamefont {C.}~\bibnamefont {Bechinger}},\ }\bibfield  {title} {\bibinfo
  {title} {Microswimmers in patterned environments},\ }\href
  {https://doi.org/10.1039/c1sm05960b} {\bibfield  {journal} {\bibinfo
  {journal} {Soft Matter}\ }\textbf {\bibinfo {volume} {7}},\ \bibinfo {pages}
  {8810} (\bibinfo {year} {2011})}\BibitemShut {NoStop}%
\bibitem [{\citenamefont {Halder}\ and\ \citenamefont
  {Khan}(2025)}]{Halder2025}%
  \BibitemOpen
  \bibfield  {author} {\bibinfo {author} {\bibfnamefont {S.}~\bibnamefont
  {Halder}}\ and\ \bibinfo {author} {\bibfnamefont {M.}~\bibnamefont {Khan}},\
  }\href {https://doi.org/10.48550/ARXIV.2505.07265} {\bibinfo {title}
  {Interplay between timescales governs the residual activity of a harmonically
  bound active brownian particle}} (\bibinfo {year} {2025})\BibitemShut
  {NoStop}%
\bibitem [{\citenamefont {Romanczuk}\ \emph {et~al.}(2012)\citenamefont
  {Romanczuk}, \citenamefont {Bär}, \citenamefont {Ebeling}, \citenamefont
  {Lindner},\ and\ \citenamefont {Schimansky-Geier}}]{Romanczuk2012}%
  \BibitemOpen
  \bibfield  {author} {\bibinfo {author} {\bibfnamefont {P.}~\bibnamefont
  {Romanczuk}}, \bibinfo {author} {\bibfnamefont {M.}~\bibnamefont {Bär}},
  \bibinfo {author} {\bibfnamefont {W.}~\bibnamefont {Ebeling}}, \bibinfo
  {author} {\bibfnamefont {B.}~\bibnamefont {Lindner}},\ and\ \bibinfo {author}
  {\bibfnamefont {L.}~\bibnamefont {Schimansky-Geier}},\ }\bibfield  {title}
  {\bibinfo {title} {Active brownian particles: From individual to collective
  stochastic dynamics},\ }\href {https://doi.org/10.1140/epjst/e2012-01529-y}
  {\bibfield  {journal} {\bibinfo  {journal} {The European Physical Journal
  Special Topics}\ }\textbf {\bibinfo {volume} {202}},\ \bibinfo {pages} {1}
  (\bibinfo {year} {2012})}\BibitemShut {NoStop}%
\bibitem [{\citenamefont {Fodor}\ and\ \citenamefont
  {Cristina Marchetti}(2018)}]{Fodor2018}%
  \BibitemOpen
  \bibfield  {author} {\bibinfo {author} {\bibfnamefont {E.}~\bibnamefont
  {Fodor}}\ and\ \bibinfo {author} {\bibfnamefont {M.}~\bibnamefont
  {Cristina Marchetti}},\ }\bibfield  {title} {\bibinfo {title} {The
  statistical physics of active matter: From self-catalytic colloids to living
  cells},\ }\href {https://doi.org/10.1016/j.physa.2017.12.137} {\bibfield
  {journal} {\bibinfo  {journal} {Physica A: Statistical Mechanics and its
  Applications}\ }\textbf {\bibinfo {volume} {504}},\ \bibinfo {pages} {106}
  (\bibinfo {year} {2018})}\BibitemShut {NoStop}%
\bibitem [{\citenamefont {Basu}\ \emph {et~al.}(2018)\citenamefont {Basu},
  \citenamefont {Majumdar}, \citenamefont {Rosso},\ and\ \citenamefont
  {Schehr}}]{Basu2018}%
  \BibitemOpen
  \bibfield  {author} {\bibinfo {author} {\bibfnamefont {U.}~\bibnamefont
  {Basu}}, \bibinfo {author} {\bibfnamefont {S.~N.}\ \bibnamefont {Majumdar}},
  \bibinfo {author} {\bibfnamefont {A.}~\bibnamefont {Rosso}},\ and\ \bibinfo
  {author} {\bibfnamefont {G.}~\bibnamefont {Schehr}},\ }\bibfield  {title}
  {\bibinfo {title} {Active brownian motion in two dimensions},\ }\href
  {https://doi.org/10.1103/physreve.98.062121} {\bibfield  {journal} {\bibinfo
  {journal} {Physical Review E}\ }\textbf {\bibinfo {volume} {98}},\ \bibinfo
  {pages} {062121} (\bibinfo {year} {2018})}\BibitemShut {NoStop}%
\bibitem [{\citenamefont {Cates}\ and\ \citenamefont
  {Tailleur}(2013)}]{Cates2013}%
  \BibitemOpen
  \bibfield  {author} {\bibinfo {author} {\bibfnamefont {M.~E.}\ \bibnamefont
  {Cates}}\ and\ \bibinfo {author} {\bibfnamefont {J.}~\bibnamefont
  {Tailleur}},\ }\bibfield  {title} {\bibinfo {title} {When are active brownian
  particles and run-and-tumble particles equivalent? consequences for
  motility-induced phase separation},\ }\href
  {https://doi.org/10.1209/0295-5075/101/20010} {\bibfield  {journal} {\bibinfo
   {journal} {EPL (Europhysics Letters)}\ }\textbf {\bibinfo {volume} {101}},\
  \bibinfo {pages} {20010} (\bibinfo {year} {2013})}\BibitemShut {NoStop}%
\bibitem [{\citenamefont {Solon}\ \emph {et~al.}(2015)\citenamefont {Solon},
  \citenamefont {Cates},\ and\ \citenamefont {Tailleur}}]{Solon2015}%
  \BibitemOpen
  \bibfield  {author} {\bibinfo {author} {\bibfnamefont {A.~P.}\ \bibnamefont
  {Solon}}, \bibinfo {author} {\bibfnamefont {M.~E.}\ \bibnamefont {Cates}},\
  and\ \bibinfo {author} {\bibfnamefont {J.}~\bibnamefont {Tailleur}},\
  }\bibfield  {title} {\bibinfo {title} {Active brownian particles and
  run-and-tumble particles: A comparative study},\ }\href
  {https://doi.org/10.1140/epjst/e2015-02457-0} {\bibfield  {journal} {\bibinfo
   {journal} {The European Physical Journal Special Topics}\ }\textbf {\bibinfo
  {volume} {224}},\ \bibinfo {pages} {1231} (\bibinfo {year}
  {2015})}\BibitemShut {NoStop}%
\bibitem [{\citenamefont {Khatami}\ \emph {et~al.}(2016)\citenamefont
  {Khatami}, \citenamefont {Wolff}, \citenamefont {Pohl}, \citenamefont
  {Ejtehadi},\ and\ \citenamefont {Stark}}]{Khatami2016}%
  \BibitemOpen
  \bibfield  {author} {\bibinfo {author} {\bibfnamefont {M.}~\bibnamefont
  {Khatami}}, \bibinfo {author} {\bibfnamefont {K.}~\bibnamefont {Wolff}},
  \bibinfo {author} {\bibfnamefont {O.}~\bibnamefont {Pohl}}, \bibinfo {author}
  {\bibfnamefont {M.~R.}\ \bibnamefont {Ejtehadi}},\ and\ \bibinfo {author}
  {\bibfnamefont {H.}~\bibnamefont {Stark}},\ }\bibfield  {title} {\bibinfo
  {title} {Active brownian particles and run-and-tumble particles separate
  inside a maze},\ }\bibfield  {journal} {\bibinfo  {journal} {Scientific
  Reports}\ }\textbf {\bibinfo {volume} {6}},\ \href
  {https://doi.org/10.1038/srep37670} {10.1038/srep37670} (\bibinfo {year}
  {2016})\BibitemShut {NoStop}%
\bibitem [{\citenamefont {Matthäus}\ \emph {et~al.}(2009)\citenamefont
  {Matthäus}, \citenamefont {Jagodič},\ and\ \citenamefont
  {Dobnikar}}]{Matthaeus2009}%
  \BibitemOpen
  \bibfield  {author} {\bibinfo {author} {\bibfnamefont {F.}~\bibnamefont
  {Matthäus}}, \bibinfo {author} {\bibfnamefont {M.}~\bibnamefont
  {Jagodič}},\ and\ \bibinfo {author} {\bibfnamefont {J.}~\bibnamefont
  {Dobnikar}},\ }\bibfield  {title} {\bibinfo {title} {E. coli superdiffusion
  and chemotaxis—search strategy, precision, and motility},\ }\href
  {https://doi.org/10.1016/j.bpj.2009.04.065} {\bibfield  {journal} {\bibinfo
  {journal} {Biophysical Journal}\ }\textbf {\bibinfo {volume} {97}},\ \bibinfo
  {pages} {946} (\bibinfo {year} {2009})}\BibitemShut {NoStop}%
\bibitem [{\citenamefont {Zaburdaev}\ \emph {et~al.}(2015)\citenamefont
  {Zaburdaev}, \citenamefont {Denisov},\ and\ \citenamefont
  {Klafter}}]{Zaburdaev2015}%
  \BibitemOpen
  \bibfield  {author} {\bibinfo {author} {\bibfnamefont {V.}~\bibnamefont
  {Zaburdaev}}, \bibinfo {author} {\bibfnamefont {S.}~\bibnamefont {Denisov}},\
  and\ \bibinfo {author} {\bibfnamefont {J.}~\bibnamefont {Klafter}},\
  }\bibfield  {title} {\bibinfo {title} {Lévy walks},\ }\href
  {https://doi.org/10.1103/revmodphys.87.483} {\bibfield  {journal} {\bibinfo
  {journal} {Reviews of Modern Physics}\ }\textbf {\bibinfo {volume} {87}},\
  \bibinfo {pages} {483} (\bibinfo {year} {2015})}\BibitemShut {NoStop}%
\bibitem [{\citenamefont {Rupprecht}\ \emph {et~al.}(2016)\citenamefont
  {Rupprecht}, \citenamefont {Bénichou},\ and\ \citenamefont
  {Voituriez}}]{Rupprecht2016}%
  \BibitemOpen
  \bibfield  {author} {\bibinfo {author} {\bibfnamefont {J.-F.}\ \bibnamefont
  {Rupprecht}}, \bibinfo {author} {\bibfnamefont {O.}~\bibnamefont
  {Bénichou}},\ and\ \bibinfo {author} {\bibfnamefont {R.}~\bibnamefont
  {Voituriez}},\ }\bibfield  {title} {\bibinfo {title} {Optimal search
  strategies of run-and-tumble walks},\ }\href
  {https://doi.org/10.1103/physreve.94.012117} {\bibfield  {journal} {\bibinfo
  {journal} {Physical Review E}\ }\textbf {\bibinfo {volume} {94}},\ \bibinfo
  {pages} {012117} (\bibinfo {year} {2016})}\BibitemShut {NoStop}%
\bibitem [{\citenamefont {Kirkegaard}\ and\ \citenamefont
  {Goldstein}(2018)}]{Kirkegaard2018}%
  \BibitemOpen
  \bibfield  {author} {\bibinfo {author} {\bibfnamefont {J.~B.}\ \bibnamefont
  {Kirkegaard}}\ and\ \bibinfo {author} {\bibfnamefont {R.~E.}\ \bibnamefont
  {Goldstein}},\ }\bibfield  {title} {\bibinfo {title} {The role of tumbling
  frequency and persistence in optimal run-and-tumble chemotaxis},\ }\href
  {https://doi.org/10.1093/imamat/hxy013} {\bibfield  {journal} {\bibinfo
  {journal} {IMA Journal of Applied Mathematics}\ }\textbf {\bibinfo {volume}
  {83}},\ \bibinfo {pages} {700} (\bibinfo {year} {2018})}\BibitemShut
  {NoStop}%
\bibitem [{\citenamefont {Nash}\ \emph {et~al.}(2010)\citenamefont {Nash},
  \citenamefont {Adhikari}, \citenamefont {Tailleur},\ and\ \citenamefont
  {Cates}}]{Nash2010}%
  \BibitemOpen
  \bibfield  {author} {\bibinfo {author} {\bibfnamefont {R.~W.}\ \bibnamefont
  {Nash}}, \bibinfo {author} {\bibfnamefont {R.}~\bibnamefont {Adhikari}},
  \bibinfo {author} {\bibfnamefont {J.}~\bibnamefont {Tailleur}},\ and\
  \bibinfo {author} {\bibfnamefont {M.~E.}\ \bibnamefont {Cates}},\ }\bibfield
  {title} {\bibinfo {title} {Run-and-tumble particles with hydrodynamics:
  Sedimentation, trapping, and upstream swimming},\ }\href
  {https://doi.org/10.1103/physrevlett.104.258101} {\bibfield  {journal}
  {\bibinfo  {journal} {Physical Review Letters}\ }\textbf {\bibinfo {volume}
  {104}},\ \bibinfo {pages} {258101} (\bibinfo {year} {2010})}\BibitemShut
  {NoStop}%
\bibitem [{\citenamefont {Santra}\ \emph {et~al.}(2020)\citenamefont {Santra},
  \citenamefont {Basu},\ and\ \citenamefont {Sabhapandit}}]{Santra2020}%
  \BibitemOpen
  \bibfield  {author} {\bibinfo {author} {\bibfnamefont {I.}~\bibnamefont
  {Santra}}, \bibinfo {author} {\bibfnamefont {U.}~\bibnamefont {Basu}},\ and\
  \bibinfo {author} {\bibfnamefont {S.}~\bibnamefont {Sabhapandit}},\
  }\bibfield  {title} {\bibinfo {title} {Run-and-tumble particles in two
  dimensions: Marginal position distributions},\ }\href
  {https://doi.org/10.1103/physreve.101.062120} {\bibfield  {journal} {\bibinfo
   {journal} {Physical Review E}\ }\textbf {\bibinfo {volume} {101}},\ \bibinfo
  {pages} {062120} (\bibinfo {year} {2020})}\BibitemShut {NoStop}%
\bibitem [{\citenamefont {Mallikarjun}\ and\ \citenamefont
  {Pal}(2023)}]{mallikarjun2023chiral}%
  \BibitemOpen
  \bibfield  {author} {\bibinfo {author} {\bibfnamefont {R.}~\bibnamefont
  {Mallikarjun}}\ and\ \bibinfo {author} {\bibfnamefont {A.}~\bibnamefont
  {Pal}},\ }\bibfield  {title} {\bibinfo {title} {Chiral run-and-tumble walker:
  Transport and optimizing search},\ }\href
  {https://doi.org/10.1016/j.physa.2023.128821} {\bibfield  {journal} {\bibinfo
   {journal} {Physica A: Statistical Mechanics and its Applications}\ }\textbf
  {\bibinfo {volume} {622}},\ \bibinfo {pages} {128821} (\bibinfo {year}
  {2023})}\BibitemShut {NoStop}%
\bibitem [{\citenamefont {Zhang}\ and\ \citenamefont
  {Pruessner}(2022)}]{Zhang2022}%
  \BibitemOpen
  \bibfield  {author} {\bibinfo {author} {\bibfnamefont {Z.}~\bibnamefont
  {Zhang}}\ and\ \bibinfo {author} {\bibfnamefont {G.}~\bibnamefont
  {Pruessner}},\ }\bibfield  {title} {\bibinfo {title} {Field theory of free
  run and tumble particles in d dimensions},\ }\href
  {https://doi.org/10.1088/1751-8121/ac37e6} {\bibfield  {journal} {\bibinfo
  {journal} {Journal of Physics A: Mathematical and Theoretical}\ }\textbf
  {\bibinfo {volume} {55}},\ \bibinfo {pages} {045204} (\bibinfo {year}
  {2022})}\BibitemShut {NoStop}%
\bibitem [{\citenamefont {Huo}\ \emph {et~al.}(2021)\citenamefont {Huo},
  \citenamefont {He}, \citenamefont {Zhang},\ and\ \citenamefont
  {Yuan}}]{Huo2021}%
  \BibitemOpen
  \bibfield  {author} {\bibinfo {author} {\bibfnamefont {H.}~\bibnamefont
  {Huo}}, \bibinfo {author} {\bibfnamefont {R.}~\bibnamefont {He}}, \bibinfo
  {author} {\bibfnamefont {R.}~\bibnamefont {Zhang}},\ and\ \bibinfo {author}
  {\bibfnamefont {J.}~\bibnamefont {Yuan}},\ }\bibfield  {title} {\bibinfo
  {title} {Swimming escherichia coli cells explore the environment by lévy
  walk},\ }\bibfield  {journal} {\bibinfo  {journal} {Applied and Environmental
  Microbiology}\ }\textbf {\bibinfo {volume} {87}},\ \href
  {https://doi.org/10.1128/aem.02429-20} {10.1128/aem.02429-20} (\bibinfo
  {year} {2021})\BibitemShut {NoStop}%
\bibitem [{\citenamefont {Saragosti}\ \emph {et~al.}(2011)\citenamefont
  {Saragosti}, \citenamefont {Calvez}, \citenamefont {Bournaveas},
  \citenamefont {Perthame}, \citenamefont {Buguin},\ and\ \citenamefont
  {Silberzan}}]{Saragosti2011}%
  \BibitemOpen
  \bibfield  {author} {\bibinfo {author} {\bibfnamefont {J.}~\bibnamefont
  {Saragosti}}, \bibinfo {author} {\bibfnamefont {V.}~\bibnamefont {Calvez}},
  \bibinfo {author} {\bibfnamefont {N.}~\bibnamefont {Bournaveas}}, \bibinfo
  {author} {\bibfnamefont {B.}~\bibnamefont {Perthame}}, \bibinfo {author}
  {\bibfnamefont {A.}~\bibnamefont {Buguin}},\ and\ \bibinfo {author}
  {\bibfnamefont {P.}~\bibnamefont {Silberzan}},\ }\bibfield  {title} {\bibinfo
  {title} {Directional persistence of chemotactic bacteria in a traveling
  concentration wave},\ }\href {https://doi.org/10.1073/pnas.1101996108}
  {\bibfield  {journal} {\bibinfo  {journal} {Proceedings of the National
  Academy of Sciences}\ }\textbf {\bibinfo {volume} {108}},\ \bibinfo {pages}
  {16235} (\bibinfo {year} {2011})}\BibitemShut {NoStop}%
\bibitem [{\citenamefont {Taktikos}\ \emph {et~al.}(2013)\citenamefont
  {Taktikos}, \citenamefont {Stark},\ and\ \citenamefont
  {Zaburdaev}}]{Taktikos2013}%
  \BibitemOpen
  \bibfield  {author} {\bibinfo {author} {\bibfnamefont {J.}~\bibnamefont
  {Taktikos}}, \bibinfo {author} {\bibfnamefont {H.}~\bibnamefont {Stark}},\
  and\ \bibinfo {author} {\bibfnamefont {V.}~\bibnamefont {Zaburdaev}},\
  }\bibfield  {title} {\bibinfo {title} {How the motility pattern of bacteria
  affects their dispersal and chemotaxis},\ }\href
  {https://doi.org/10.1371/journal.pone.0081936} {\bibfield  {journal}
  {\bibinfo  {journal} {PLoS ONE}\ }\textbf {\bibinfo {volume} {8}},\ \bibinfo
  {pages} {e81936} (\bibinfo {year} {2013})}\BibitemShut {NoStop}%
\bibitem [{\citenamefont {Colin}\ \emph {et~al.}(2021)\citenamefont {Colin},
  \citenamefont {Ni}, \citenamefont {Laganenka},\ and\ \citenamefont
  {Sourjik}}]{Colin2021}%
  \BibitemOpen
  \bibfield  {author} {\bibinfo {author} {\bibfnamefont {R.}~\bibnamefont
  {Colin}}, \bibinfo {author} {\bibfnamefont {B.}~\bibnamefont {Ni}}, \bibinfo
  {author} {\bibfnamefont {L.}~\bibnamefont {Laganenka}},\ and\ \bibinfo
  {author} {\bibfnamefont {V.}~\bibnamefont {Sourjik}},\ }\bibfield  {title}
  {\bibinfo {title} {Multiple functions of flagellar motility and chemotaxis in
  bacterial physiology},\ }\bibfield  {journal} {\bibinfo  {journal} {FEMS
  Microbiology Reviews}\ }\textbf {\bibinfo {volume} {45}},\ \href
  {https://doi.org/10.1093/femsre/fuab038} {10.1093/femsre/fuab038} (\bibinfo
  {year} {2021})\BibitemShut {NoStop}%
\bibitem [{\citenamefont {Raza}\ \emph {et~al.}(2023)\citenamefont {Raza},
  \citenamefont {George}, \citenamefont {Kumari}, \citenamefont {Mitra},\ and\
  \citenamefont {Paul}}]{Raza2023}%
  \BibitemOpen
  \bibfield  {author} {\bibinfo {author} {\bibfnamefont {M.~R.}\ \bibnamefont
  {Raza}}, \bibinfo {author} {\bibfnamefont {J.~E.}\ \bibnamefont {George}},
  \bibinfo {author} {\bibfnamefont {S.}~\bibnamefont {Kumari}}, \bibinfo
  {author} {\bibfnamefont {M.~K.}\ \bibnamefont {Mitra}},\ and\ \bibinfo
  {author} {\bibfnamefont {D.}~\bibnamefont {Paul}},\ }\bibfield  {title}
  {\bibinfo {title} {Anomalous diffusion of e. coli under microfluidic
  confinement and chemical gradient},\ }\href
  {https://doi.org/10.1039/d3sm00286a} {\bibfield  {journal} {\bibinfo
  {journal} {Soft Matter}\ }\textbf {\bibinfo {volume} {19}},\ \bibinfo {pages}
  {6446} (\bibinfo {year} {2023})}\BibitemShut {NoStop}%
\bibitem [{\citenamefont {Wang}\ and\ \citenamefont {Tsang}(2025)}]{Wang2025}%
  \BibitemOpen
  \bibfield  {author} {\bibinfo {author} {\bibfnamefont {Z.}~\bibnamefont
  {Wang}}\ and\ \bibinfo {author} {\bibfnamefont {A.~C.~H.}\ \bibnamefont
  {Tsang}},\ }\bibfield  {title} {\bibinfo {title} {Intermediate light
  adaptation induces oscillatory phototaxis switching and pattern formation in
  chlamydomonas},\ }\bibfield  {journal} {\bibinfo  {journal} {Proceedings of
  the National Academy of Sciences}\ }\textbf {\bibinfo {volume} {122}},\ \href
  {https://doi.org/10.1073/pnas.2425369122} {10.1073/pnas.2425369122} (\bibinfo
  {year} {2025})\BibitemShut {NoStop}%
\bibitem [{\citenamefont {Lozano}\ \emph {et~al.}(2018)\citenamefont {Lozano},
  \citenamefont {Gomez-Solano},\ and\ \citenamefont {Bechinger}}]{Lozano2018}%
  \BibitemOpen
  \bibfield  {author} {\bibinfo {author} {\bibfnamefont {C.}~\bibnamefont
  {Lozano}}, \bibinfo {author} {\bibfnamefont {J.~R.}\ \bibnamefont
  {Gomez-Solano}},\ and\ \bibinfo {author} {\bibfnamefont {C.}~\bibnamefont
  {Bechinger}},\ }\bibfield  {title} {\bibinfo {title} {Run-and-tumble-like
  motion of active colloids in viscoelastic media},\ }\href
  {https://doi.org/10.1088/1367-2630/aa9ed1} {\bibfield  {journal} {\bibinfo
  {journal} {New Journal of Physics}\ }\textbf {\bibinfo {volume} {20}},\
  \bibinfo {pages} {015008} (\bibinfo {year} {2018})}\BibitemShut {NoStop}%
\bibitem [{\citenamefont {Paramanick}\ \emph {et~al.}(2025)\citenamefont
  {Paramanick}, \citenamefont {Pardhi}, \citenamefont {Soni},\ and\
  \citenamefont {Kumar}}]{Paramanick2025}%
  \BibitemOpen
  \bibfield  {author} {\bibinfo {author} {\bibfnamefont {S.}~\bibnamefont
  {Paramanick}}, \bibinfo {author} {\bibfnamefont {U.}~\bibnamefont {Pardhi}},
  \bibinfo {author} {\bibfnamefont {H.}~\bibnamefont {Soni}},\ and\ \bibinfo
  {author} {\bibfnamefont {N.}~\bibnamefont {Kumar}},\ }\href
  {https://doi.org/10.48550/ARXIV.2502.01257} {\bibinfo {title} {Spontaneous
  emergence of run-and-tumble-like dynamics in coupled self-propelled robots:
  experiment and theory}} (\bibinfo {year} {2025})\BibitemShut {NoStop}%
\bibitem [{\citenamefont {Kumar}\ \emph {et~al.}(2020)\citenamefont {Kumar},
  \citenamefont {Sadekar},\ and\ \citenamefont {Basu}}]{Kumar2020}%
  \BibitemOpen
  \bibfield  {author} {\bibinfo {author} {\bibfnamefont {V.}~\bibnamefont
  {Kumar}}, \bibinfo {author} {\bibfnamefont {O.}~\bibnamefont {Sadekar}},\
  and\ \bibinfo {author} {\bibfnamefont {U.}~\bibnamefont {Basu}},\ }\bibfield
  {title} {\bibinfo {title} {Active brownian motion in two dimensions under
  stochastic resetting},\ }\href {https://doi.org/10.1103/physreve.102.052129}
  {\bibfield  {journal} {\bibinfo  {journal} {Physical Review E}\ }\textbf
  {\bibinfo {volume} {102}},\ \bibinfo {pages} {052129} (\bibinfo {year}
  {2020})}\BibitemShut {NoStop}%
\bibitem [{\citenamefont {Sar}\ \emph {et~al.}(2023)\citenamefont {Sar},
  \citenamefont {Ray}, \citenamefont {Ghosh}, \citenamefont {Hens},\ and\
  \citenamefont {Pal}}]{sar2023resetting}%
  \BibitemOpen
  \bibfield  {author} {\bibinfo {author} {\bibfnamefont {G.~K.}\ \bibnamefont
  {Sar}}, \bibinfo {author} {\bibfnamefont {A.}~\bibnamefont {Ray}}, \bibinfo
  {author} {\bibfnamefont {D.}~\bibnamefont {Ghosh}}, \bibinfo {author}
  {\bibfnamefont {C.}~\bibnamefont {Hens}},\ and\ \bibinfo {author}
  {\bibfnamefont {A.}~\bibnamefont {Pal}},\ }\bibfield  {title} {\bibinfo
  {title} {Resetting-mediated navigation of an active brownian searcher in a
  homogeneous topography},\ }\href {https://doi.org/10.1039/D3SM00271C}
  {\bibfield  {journal} {\bibinfo  {journal} {Soft Matter}\ }\textbf {\bibinfo
  {volume} {19}},\ \bibinfo {pages} {4502} (\bibinfo {year}
  {2023})}\BibitemShut {NoStop}%
\bibitem [{\citenamefont {Baouche}\ \emph {et~al.}(2024)\citenamefont
  {Baouche}, \citenamefont {Franosch}, \citenamefont {Meiners},\ and\
  \citenamefont {Kurzthaler}}]{Baouche2024}%
  \BibitemOpen
  \bibfield  {author} {\bibinfo {author} {\bibfnamefont {Y.}~\bibnamefont
  {Baouche}}, \bibinfo {author} {\bibfnamefont {T.}~\bibnamefont {Franosch}},
  \bibinfo {author} {\bibfnamefont {M.}~\bibnamefont {Meiners}},\ and\ \bibinfo
  {author} {\bibfnamefont {C.}~\bibnamefont {Kurzthaler}},\ }\bibfield  {title}
  {\bibinfo {title} {Active brownian particle under stochastic orientational
  resetting},\ }\href {https://doi.org/10.1088/1367-2630/ad602a} {\bibfield
  {journal} {\bibinfo  {journal} {New Journal of Physics}\ }\textbf {\bibinfo
  {volume} {26}},\ \bibinfo {pages} {073041} (\bibinfo {year}
  {2024})}\BibitemShut {NoStop}%
\bibitem [{\citenamefont {Paramanick}\ \emph
  {et~al.}(2024{\natexlab{a}})\citenamefont {Paramanick}, \citenamefont
  {Biswas}, \citenamefont {Soni}, \citenamefont {Pal},\ and\ \citenamefont
  {Kumar}}]{Paramanick2024a}%
  \BibitemOpen
  \bibfield  {author} {\bibinfo {author} {\bibfnamefont {S.}~\bibnamefont
  {Paramanick}}, \bibinfo {author} {\bibfnamefont {A.}~\bibnamefont {Biswas}},
  \bibinfo {author} {\bibfnamefont {H.}~\bibnamefont {Soni}}, \bibinfo {author}
  {\bibfnamefont {A.}~\bibnamefont {Pal}},\ and\ \bibinfo {author}
  {\bibfnamefont {N.}~\bibnamefont {Kumar}},\ }\bibfield  {title} {\bibinfo
  {title} {Uncovering universal characteristics of homing paths using foraging
  robots},\ }\href {https://doi.org/10.1103/prxlife.2.033007} {\bibfield
  {journal} {\bibinfo  {journal} {PRX Life}\ }\textbf {\bibinfo {volume} {2}},\
  \bibinfo {pages} {033007} (\bibinfo {year} {2024}{\natexlab{a}})}\BibitemShut
  {NoStop}%
\bibitem [{\citenamefont {Baouche}\ and\ \citenamefont
  {Kurzthaler}(2025)}]{Baouche2025}%
  \BibitemOpen
  \bibfield  {author} {\bibinfo {author} {\bibfnamefont {Y.}~\bibnamefont
  {Baouche}}\ and\ \bibinfo {author} {\bibfnamefont {C.}~\bibnamefont
  {Kurzthaler}},\ }\bibfield  {title} {\bibinfo {title} {Optimal first-passage
  times of active brownian particles under stochastic resetting},\ }\href
  {https://doi.org/10.1039/d5sm00340g} {\bibfield  {journal} {\bibinfo
  {journal} {Soft Matter}\ }\textbf {\bibinfo {volume} {21}},\ \bibinfo {pages}
  {5998} (\bibinfo {year} {2025})}\BibitemShut {NoStop}%
\bibitem [{\citenamefont {Shee}(2025)}]{Shee2025}%
  \BibitemOpen
  \bibfield  {author} {\bibinfo {author} {\bibfnamefont {A.}~\bibnamefont
  {Shee}},\ }\bibfield  {title} {\bibinfo {title} {Active brownian particle
  under stochastic position and orientation resetting in a harmonic trap},\
  }\href {https://doi.org/10.1088/2399-6528/adb36e} {\bibfield  {journal}
  {\bibinfo  {journal} {Journal of Physics Communications}\ }\textbf {\bibinfo
  {volume} {9}},\ \bibinfo {pages} {025003} (\bibinfo {year}
  {2025})}\BibitemShut {NoStop}%
\bibitem [{\citenamefont {Lee}\ \emph {et~al.}(2019)\citenamefont {Lee},
  \citenamefont {Szuttor},\ and\ \citenamefont {Holm}}]{Lee2019}%
  \BibitemOpen
  \bibfield  {author} {\bibinfo {author} {\bibfnamefont {M.}~\bibnamefont
  {Lee}}, \bibinfo {author} {\bibfnamefont {K.}~\bibnamefont {Szuttor}},\ and\
  \bibinfo {author} {\bibfnamefont {C.}~\bibnamefont {Holm}},\ }\bibfield
  {title} {\bibinfo {title} {A computational model for bacterial run-and-tumble
  motion},\ }\bibfield  {journal} {\bibinfo  {journal} {The Journal of Chemical
  Physics}\ }\textbf {\bibinfo {volume} {150}},\ \href
  {https://doi.org/10.1063/1.5085836} {10.1063/1.5085836} (\bibinfo {year}
  {2019})\BibitemShut {NoStop}%
\bibitem [{\citenamefont {Evans}\ and\ \citenamefont
  {Majumdar}(2011)}]{evans_diffusion_2011}%
  \BibitemOpen
  \bibfield  {author} {\bibinfo {author} {\bibfnamefont {M.~R.}\ \bibnamefont
  {Evans}}\ and\ \bibinfo {author} {\bibfnamefont {S.~N.}\ \bibnamefont
  {Majumdar}},\ }\bibfield  {title} {\bibinfo {title} {Diffusion with
  {Stochastic} {Resetting}},\ }\href
  {https://doi.org/10.1103/PhysRevLett.106.160601} {\bibfield  {journal}
  {\bibinfo  {journal} {Physical Review Letters}\ }\textbf {\bibinfo {volume}
  {106}},\ \bibinfo {pages} {160601} (\bibinfo {year} {2011})}\BibitemShut
  {NoStop}%
\bibitem [{\citenamefont {Evans}\ \emph {et~al.}(2020)\citenamefont {Evans},
  \citenamefont {Majumdar},\ and\ \citenamefont
  {Schehr}}]{evans_stochastic_2020}%
  \BibitemOpen
  \bibfield  {author} {\bibinfo {author} {\bibfnamefont {M.~R.}\ \bibnamefont
  {Evans}}, \bibinfo {author} {\bibfnamefont {S.~N.}\ \bibnamefont
  {Majumdar}},\ and\ \bibinfo {author} {\bibfnamefont {G.}~\bibnamefont
  {Schehr}},\ }\bibfield  {title} {\bibinfo {title} {Stochastic resetting and
  applications},\ }\href {https://doi.org/10.1088/1751-8121/ab7cfe} {\bibfield
  {journal} {\bibinfo  {journal} {Journal of Physics A: Mathematical and
  Theoretical}\ }\textbf {\bibinfo {volume} {53}},\ \bibinfo {pages} {193001}
  (\bibinfo {year} {2020})}\BibitemShut {NoStop}%
\bibitem [{\citenamefont {Pal}\ \emph {et~al.}(2024)\citenamefont {Pal},
  \citenamefont {Stojkoski},\ and\ \citenamefont {Sandev}}]{pal2024random}%
  \BibitemOpen
  \bibfield  {author} {\bibinfo {author} {\bibfnamefont {A.}~\bibnamefont
  {Pal}}, \bibinfo {author} {\bibfnamefont {V.}~\bibnamefont {Stojkoski}},\
  and\ \bibinfo {author} {\bibfnamefont {T.}~\bibnamefont {Sandev}},\
  }\bibfield  {title} {\bibinfo {title} {Random resetting in search problems},\
  }in\ \href@noop {} {\emph {\bibinfo {booktitle} {Target Search Problems}}}\
  (\bibinfo  {publisher} {Springer},\ \bibinfo {year} {2024})\ pp.\ \bibinfo
  {pages} {323--355}\BibitemShut {NoStop}%
\bibitem [{\citenamefont {Tal-Friedman}\ \emph {et~al.}(2020)\citenamefont
  {Tal-Friedman}, \citenamefont {Pal}, \citenamefont {Sekhon}, \citenamefont
  {Reuveni},\ and\ \citenamefont {Roichman}}]{tal2020experimental}%
  \BibitemOpen
  \bibfield  {author} {\bibinfo {author} {\bibfnamefont {O.}~\bibnamefont
  {Tal-Friedman}}, \bibinfo {author} {\bibfnamefont {A.}~\bibnamefont {Pal}},
  \bibinfo {author} {\bibfnamefont {A.}~\bibnamefont {Sekhon}}, \bibinfo
  {author} {\bibfnamefont {S.}~\bibnamefont {Reuveni}},\ and\ \bibinfo {author}
  {\bibfnamefont {Y.}~\bibnamefont {Roichman}},\ }\bibfield  {title} {\bibinfo
  {title} {Experimental realization of diffusion with stochastic resetting},\
  }\href {https://doi.org/10.1021/acs.jpclett.0c02122.} {\bibfield  {journal}
  {\bibinfo  {journal} {The Journal of Physical Chemistry Letters}\ }\textbf
  {\bibinfo {volume} {11}},\ \bibinfo {pages} {7350} (\bibinfo {year}
  {2020})}\BibitemShut {NoStop}%
\bibitem [{\citenamefont {Besga}\ \emph {et~al.}(2020)\citenamefont {Besga},
  \citenamefont {Bovon}, \citenamefont {Petrosyan}, \citenamefont {Majumdar},\
  and\ \citenamefont {Ciliberto}}]{besga2020optimal}%
  \BibitemOpen
  \bibfield  {author} {\bibinfo {author} {\bibfnamefont {B.}~\bibnamefont
  {Besga}}, \bibinfo {author} {\bibfnamefont {A.}~\bibnamefont {Bovon}},
  \bibinfo {author} {\bibfnamefont {A.}~\bibnamefont {Petrosyan}}, \bibinfo
  {author} {\bibfnamefont {S.~N.}\ \bibnamefont {Majumdar}},\ and\ \bibinfo
  {author} {\bibfnamefont {S.}~\bibnamefont {Ciliberto}},\ }\bibfield  {title}
  {\bibinfo {title} {Optimal mean first-passage time for a brownian searcher
  subjected to resetting: experimental and theoretical results},\ }\href
  {https://doi.org/10.1103/PhysRevResearch.2.032029} {\bibfield  {journal}
  {\bibinfo  {journal} {Physical Review Research}\ }\textbf {\bibinfo {volume}
  {2}},\ \bibinfo {pages} {032029} (\bibinfo {year} {2020})}\BibitemShut
  {NoStop}%
\bibitem [{\citenamefont {Martin}\ and\ \citenamefont
  {Snezhko}(2013)}]{Martin2013}%
  \BibitemOpen
  \bibfield  {author} {\bibinfo {author} {\bibfnamefont {J.~E.}\ \bibnamefont
  {Martin}}\ and\ \bibinfo {author} {\bibfnamefont {A.}~\bibnamefont
  {Snezhko}},\ }\bibfield  {title} {\bibinfo {title} {Driving self-assembly and
  emergent dynamics in colloidal suspensions by time-dependent magnetic
  fields},\ }\href {https://doi.org/10.1088/0034-4885/76/12/126601} {\bibfield
  {journal} {\bibinfo  {journal} {Reports on Progress in Physics}\ }\textbf
  {\bibinfo {volume} {76}},\ \bibinfo {pages} {126601} (\bibinfo {year}
  {2013})}\BibitemShut {NoStop}%
\bibitem [{\citenamefont {Han}\ \emph {et~al.}(2020)\citenamefont {Han},
  \citenamefont {Kokot}, \citenamefont {Tovkach}, \citenamefont {Glatz},
  \citenamefont {Aranson},\ and\ \citenamefont {Snezhko}}]{Han2020}%
  \BibitemOpen
  \bibfield  {author} {\bibinfo {author} {\bibfnamefont {K.}~\bibnamefont
  {Han}}, \bibinfo {author} {\bibfnamefont {G.}~\bibnamefont {Kokot}}, \bibinfo
  {author} {\bibfnamefont {O.}~\bibnamefont {Tovkach}}, \bibinfo {author}
  {\bibfnamefont {A.}~\bibnamefont {Glatz}}, \bibinfo {author} {\bibfnamefont
  {I.~S.}\ \bibnamefont {Aranson}},\ and\ \bibinfo {author} {\bibfnamefont
  {A.}~\bibnamefont {Snezhko}},\ }\bibfield  {title} {\bibinfo {title}
  {Emergence of self-organized multivortex states in flocks of active
  rollers},\ }\href {https://doi.org/10.1073/pnas.2000061117} {\bibfield
  {journal} {\bibinfo  {journal} {Proceedings of the National Academy of
  Sciences}\ }\textbf {\bibinfo {volume} {117}},\ \bibinfo {pages} {9706}
  (\bibinfo {year} {2020})}\BibitemShut {NoStop}%
\bibitem [{\citenamefont {Bishop}\ \emph {et~al.}(2023)\citenamefont {Bishop},
  \citenamefont {Biswal},\ and\ \citenamefont {Bharti}}]{Bishop2023}%
  \BibitemOpen
  \bibfield  {author} {\bibinfo {author} {\bibfnamefont {K.~J.}\ \bibnamefont
  {Bishop}}, \bibinfo {author} {\bibfnamefont {S.~L.}\ \bibnamefont {Biswal}},\
  and\ \bibinfo {author} {\bibfnamefont {B.}~\bibnamefont {Bharti}},\
  }\bibfield  {title} {\bibinfo {title} {Active colloids as models, materials,
  and machines},\ }\href
  {https://doi.org/10.1146/annurev-chembioeng-101121-084939} {\bibfield
  {journal} {\bibinfo  {journal} {Annual Review of Chemical and Biomolecular
  Engineering}\ }\textbf {\bibinfo {volume} {14}},\ \bibinfo {pages} {1}
  (\bibinfo {year} {2023})}\BibitemShut {NoStop}%
\bibitem [{\citenamefont {Ye}\ \emph {et~al.}(2023)\citenamefont {Ye},
  \citenamefont {Zhou}, \citenamefont {Zhao}, \citenamefont {Wang},
  \citenamefont {Nelson},\ and\ \citenamefont {Wang}}]{Ye2023}%
  \BibitemOpen
  \bibfield  {author} {\bibinfo {author} {\bibfnamefont {M.}~\bibnamefont
  {Ye}}, \bibinfo {author} {\bibfnamefont {Y.}~\bibnamefont {Zhou}}, \bibinfo
  {author} {\bibfnamefont {H.}~\bibnamefont {Zhao}}, \bibinfo {author}
  {\bibfnamefont {Z.}~\bibnamefont {Wang}}, \bibinfo {author} {\bibfnamefont
  {B.~J.}\ \bibnamefont {Nelson}},\ and\ \bibinfo {author} {\bibfnamefont
  {X.}~\bibnamefont {Wang}},\ }\bibfield  {title} {\bibinfo {title} {A review
  of soft microrobots: Material, fabrication, and actuation},\ }\bibfield
  {journal} {\bibinfo  {journal} {Advanced Intelligent Systems}\ }\textbf
  {\bibinfo {volume} {5}},\ \href {https://doi.org/10.1002/aisy.202300311}
  {10.1002/aisy.202300311} (\bibinfo {year} {2023})\BibitemShut {NoStop}%
\bibitem [{\citenamefont {Paramanick}\ \emph
  {et~al.}(2024{\natexlab{b}})\citenamefont {Paramanick}, \citenamefont {Pal},
  \citenamefont {Soni},\ and\ \citenamefont {Kumar}}]{Paramanick2024}%
  \BibitemOpen
  \bibfield  {author} {\bibinfo {author} {\bibfnamefont {S.}~\bibnamefont
  {Paramanick}}, \bibinfo {author} {\bibfnamefont {A.}~\bibnamefont {Pal}},
  \bibinfo {author} {\bibfnamefont {H.}~\bibnamefont {Soni}},\ and\ \bibinfo
  {author} {\bibfnamefont {N.}~\bibnamefont {Kumar}},\ }\bibfield  {title}
  {\bibinfo {title} {Programming tunable active dynamics in a self-propelled
  robot},\ }\bibfield  {journal} {\bibinfo  {journal} {The European Physical
  Journal E}\ }\textbf {\bibinfo {volume} {47}},\ \href
  {https://doi.org/10.1140/epje/s10189-024-00430-x}
  {10.1140/epje/s10189-024-00430-x} (\bibinfo {year}
  {2024}{\natexlab{b}})\BibitemShut {NoStop}%
\end{thebibliography}%


\providecommand{\noopsort}[1]{}\providecommand{\singleletter}[1]{#1}%
\begin{thebibliography}{2}%
\makeatletter
\providecommand \@ifxundefined [1]{%
 \@ifx{#1\undefined}
}%
\providecommand \@ifnum [1]{%
 \ifnum #1\expandafter \@firstoftwo
 \else \expandafter \@secondoftwo
 \fi
}%
\providecommand \@ifx [1]{%
 \ifx #1\expandafter \@firstoftwo
 \else \expandafter \@secondoftwo
 \fi
}%
\providecommand \natexlab [1]{#1}%
\providecommand \enquote  [1]{``#1''}%
\providecommand \bibnamefont  [1]{#1}%
\providecommand \bibfnamefont [1]{#1}%
\providecommand \citenamefont [1]{#1}%
\providecommand \href@noop [0]{\@secondoftwo}%
\providecommand \href [0]{\begingroup \@sanitize@url \@href}%
\providecommand \@href[1]{\@@startlink{#1}\@@href}%
\providecommand \@@href[1]{\endgroup#1\@@endlink}%
\providecommand \@sanitize@url [0]{\catcode `\\12\catcode `\$12\catcode
  `\&12\catcode `\#12\catcode `\^12\catcode `\_12\catcode `\%12\relax}%
\providecommand \@@startlink[1]{}%
\providecommand \@@endlink[0]{}%
\providecommand \url  [0]{\begingroup\@sanitize@url \@url }%
\providecommand \@url [1]{\endgroup\@href {#1}{\urlprefix }}%
\providecommand \urlprefix  [0]{URL }%
\providecommand \Eprint [0]{\href }%
\providecommand \doibase [0]{http://dx.doi.org/}%
\providecommand \selectlanguage [0]{\@gobble}%
\providecommand \bibinfo  [0]{\@secondoftwo}%
\providecommand \bibfield  [0]{\@secondoftwo}%
\providecommand \translation [1]{[#1]}%
\providecommand \BibitemOpen [0]{}%
\providecommand \bibitemStop [0]{}%
\providecommand \bibitemNoStop [0]{.\EOS\space}%
\providecommand \EOS [0]{\spacefactor3000\relax}%
\providecommand \BibitemShut  [1]{\csname bibitem#1\endcsname}%
\let\auto@bib@innerbib\@empty
\bibitem [{\citenamefont {Pal}\ \emph {et~al.}(2016)\citenamefont {Pal},
  \citenamefont {Kundu},\ and\ \citenamefont {Evans}}]{pal2016diffusion}%
  \BibitemOpen
  \bibfield  {author} {\bibinfo {author} {\bibfnamefont {A.}~\bibnamefont
  {Pal}}, \bibinfo {author} {\bibfnamefont {A.}~\bibnamefont {Kundu}}, \ and\
  \bibinfo {author} {\bibfnamefont {M.~R.}\ \bibnamefont {Evans}},\ }\href@noop
  {} {\bibfield  {journal} {\bibinfo  {journal} {Journal of Physics A:
  Mathematical and Theoretical}\ }\textbf {\bibinfo {volume} {49}},\ \bibinfo
  {pages} {225001} (\bibinfo {year} {2016})}\BibitemShut {NoStop}%
\bibitem [{\citenamefont {Kumar}\ \emph {et~al.}(2020)\citenamefont {Kumar},
  \citenamefont {Sadekar},\ and\ \citenamefont {Basu}}]{kumar2020active}%
  \BibitemOpen
  \bibfield  {author} {\bibinfo {author} {\bibfnamefont {V.}~\bibnamefont
  {Kumar}}, \bibinfo {author} {\bibfnamefont {O.}~\bibnamefont {Sadekar}}, \
  and\ \bibinfo {author} {\bibfnamefont {U.}~\bibnamefont {Basu}},\ }\href@noop
  {} {\bibfield  {journal} {\bibinfo  {journal} {Physical Review E}\ }\textbf
  {\bibinfo {volume} {102}},\ \bibinfo {pages} {052129} (\bibinfo {year}
  {2020})}\BibitemShut {NoStop}%
\end{thebibliography}%


\section{End Matter} 



\begin{figure}[htbp]
	\includegraphics[width=0.45\textwidth]{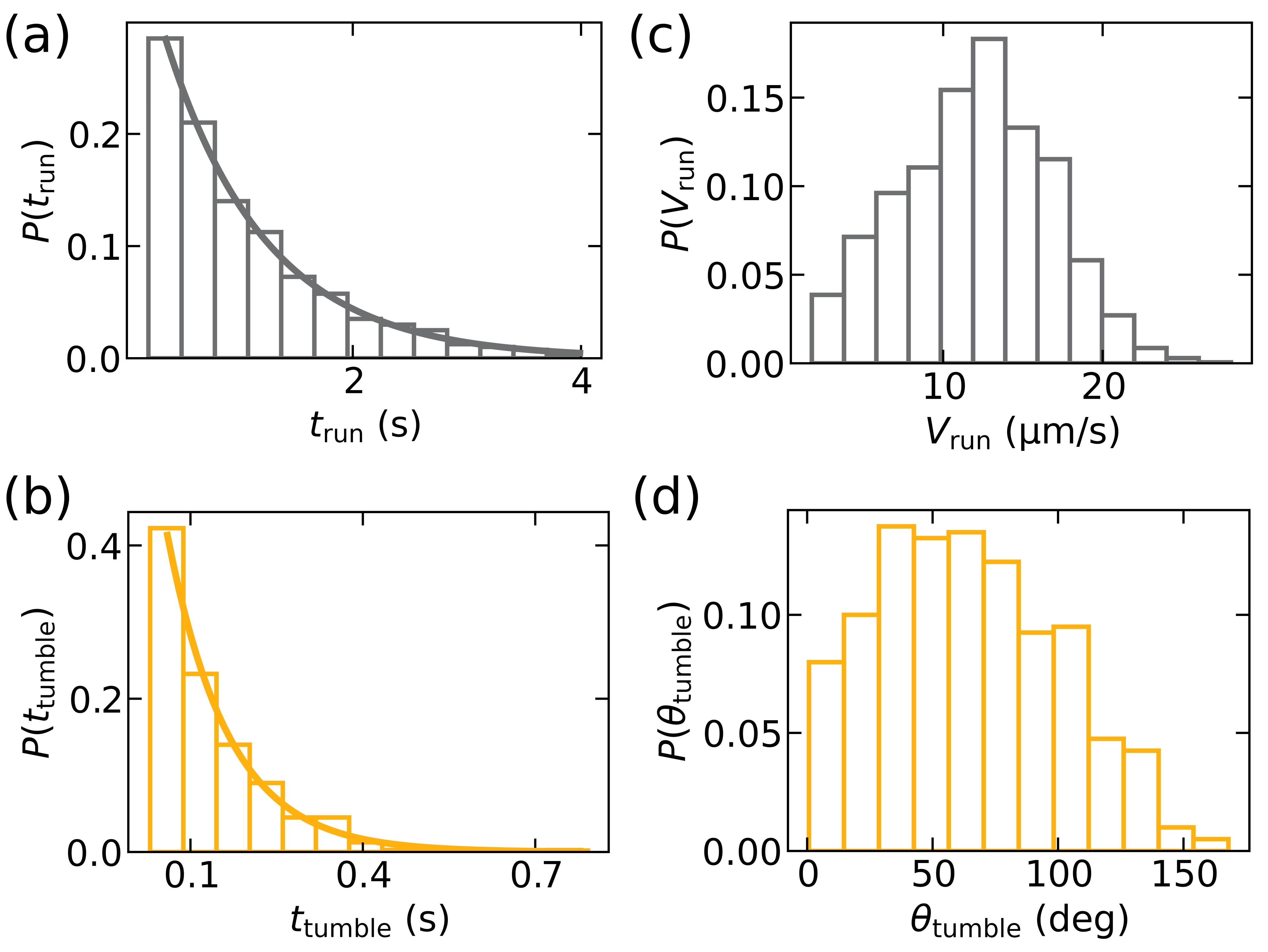}
	\caption{Probability distributions of (a) $t_{\mathrm{run}}$, (b) $t_{\mathrm{tumble}}$, (c) $V_{\mathrm{run}}$, and (d) $\theta_{\mathrm{tumble}}$ describing RnT dynamics of \textit{E. coli}. (a, b) Solid lines represent fitting to exponential distributions, providing mean values of $t_{\mathrm{run}}$ and $t_{\mathrm{tumble}}$. } 
	\label{fig:EcoliDist}
\end{figure}

\paragraph{Characterization of E.coli dynamics.} 
We cultured \textit{E. coli} (wild-type strain RP437) and studied their RnT dynamics in a dilute suspension in motility buffer (SM S1) under an inverted microscope (Nikon Ti2-U) using a 20$\times$ objective. The motions were captured at 100 fps with a CMOS camera (FLIR Grasshopper 3) attached to the microscope and tracked using the Trackpy package in Python to obtain the trajectories (Fig. \ref{fig:ABPvsRnT}(c)). The RnT dynamics was then characterized by calculating the MSD (Fig. \ref{fig:ABPvsRnT}(g)), position distribution (Fig. \ref{fig:ABPvsRnT}(h)), and probability distributions of the other relevant parameters (Fig. \ref{fig:EcoliDist}). The instantaneous swimming speed $V(t)$ and orientation $\theta (t)$ were calculated after processing the time-series data using a sliding average over three consecutive data points to remove high-frequency noise. Following the established criteria \cite{HOWARDC.1972, Berg2004}, we identified the run states when $V(t)$ continuously remained higher than \SI{7.5}{\um/\s} with a smooth variation in $\theta (t)$. On the contrary, finite-duration tumble states were recognized by a drop in $V(t)$ below \SI{7.5}{\um/\s} with a significant change in $\theta (t)$ sustained over \SI{0.06}{\second} or by a large instantaneous change in $\theta (t) >$ \ang{35} with $V(t) <$ \SI{7.5}{\um/\s} (Fig. S1). After identifying the run and tumble states, we calculated the probability distributions of run duration ($t_{\mathrm{run}}$), tumble duration ($t_{\mathrm{tumble}}$), run speed ($V_{\mathrm{run}}$), and tumble angle ($\theta_{\mathrm{tumble}}$), as shown in Fig. \ref{fig:EcoliDist}. The $t_{\mathrm{run}}$ and $t_{\mathrm{tumble}}$ distributions showed excellent fitting to exponential distributions with mean values \SI{0.88}{\second} (Fig. \ref{fig:EcoliDist}(a)) and \SI{0.11}{\second} (Fig. \ref{fig:EcoliDist}(b)), respectively; the mean $\pm$ standard deviation (SD) values for $V_{\mathrm{run}}$ and $\theta_{\mathrm{tumble}}$ were obtained as \SI[separate-uncertainty = true]{12.0(4.7)}{\um/\s} and \ang{64.8} $\pm$ \ang{38.3}, respectively.

\paragraph{Simulation of RnT dynamics.} 
We simulated RnT trajectories at time-step \SI{0.01}{\second} for a spherical shaped \textit{E.coli} of diameter (\SI{1.67}{\um}) and average swimming speed (\SI{1.02}{\um/\s}) same as those of our ABP. Considering a two-state model with finite tumble durations \cite{Lee2019} and the experimentally observed distributions of $t_{\mathrm{run}}$ and $t_{\mathrm{tumble}}$, the run-to-tumble and tumble-to-run transition rates were calculated as $1/\left\langle t_{\mathrm{run}} \right\rangle$ and $1/\left\langle t_{\mathrm{tumble}} \right\rangle$, respectively. The RnT dynamics is given by slightly modified Langevin equations compared to those for the ABP (Eq. \ref{eq:LE}): $d \bm{r} = V_{\mathrm{run}}(t) \bm{u}(\theta) dt + \sqrt{2D_{\mathrm{T}} dt} \bm{\xi}(t)$ and $d\theta = \Omega_{\mathrm{tumble}} dt + \sqrt{2D_{\mathrm{R}} dt} \eta(t)$,
where $\bm{u}(\theta)= (cos{\theta}, sin{\theta})$ is the intrinsic direction of propulsion of the RnT particles,  $\Omega_{\mathrm{tumble}}$  is the orientational speed due to tumble, and $\bm{\xi}(t) = (\xi_{x}(t),\xi_{y}(t))$, and $\eta(t)$ are independent Gaussian white noises. During the runs, the swimming direction changes slowly owing to orientational diffusion, with $\Omega_{\mathrm{tumble}}$ = 0. The run speed, $V_{\mathrm{run}}$, is considered to be zero during tumbles and remains constant during a run, but varies from one run to another. While we took $V_{\mathrm{run}}$ values from a Gaussian distribution with mean $\left\langle V_{\mathrm{run}}\right\rangle $ = \SI{1.02}{\um/\s} and a proportionately scaled down standard deviation (\SI{0.40}{\um/\s}) from that of the experimentally observed value, the $\theta_{\mathrm{tumble}}$ (= $\Omega_{\mathrm{tumble}} \times t_{\mathrm{tumble}}$) values were drawn from a Gaussian distribution with the same mean and standard deviation as those obtained experimentally.

The MSD and position distribution from the simulated RnT dynamics are shown in Fig. \ref{fig:ABPvsRnT}(i), \ref{fig:RnT}(d), and Fig. \ref{fig:ABPvsRnT}(j), \ref{fig:RnT}(e), respectively. Notably, the spherical RnT particle in the simulation experiences faster orientational diffusion during the runs than a real cylindrical-shaped \textit{E. coli} bacterium and hence exhibits slightly shorter persistence, requiring a more frequent SOR ($\lambda$ = 3) of the ABP to match the simulated RnT dynamics.

\paragraph{Characterization of ABP dynamics.}
The Pt coating (thickness of $\approx$ \SI{5}{\nm}) on one hemisphere of the Pt-silica Janus colloids acts as a catalyst for the decomposition of $\mathrm{H}_2\mathrm{O}_2$ into $\mathrm{H}_2\mathrm{O}$ and $\mathrm{O}_2$, generating a local chemical gradient and thus setting up diffusiophoretic self-propulsion \cite{Howse2007, Bechinger2016, Halder2025}. Active dynamics of the Janus microspheres were observed in a 4\% ($v/v$) aqueous suspension of $\mathrm{H}_2\mathrm{O}_2$ under an inverted microscope using a 40$\times$ objective, and brightfield images were recorded at 50 fps using a CMOS camera attached to the microscope. The Janus microspheres were tracked in the recorded image sequences using the Trackpy module in Python to obtain trajectories. A typical trajectory (Fig. S2(a)) and the corresponding ensemble- and time-averaged MSD (Fig. S2(b)) of the Janus colloids in 4\% $\mathrm{H}_2\mathrm{O}_2$ show a clear signature of activity with reference to passive dynamics in the absence of $\mathrm{H}_2\mathrm{O}_2$. The MSD exhibiting active motion was fitted to the analytical form describing ABP dynamics \cite{Howse2007, Bechinger2016, Halder2025}
\begin{equation}
	\left\langle \Delta r^2 (\tau) \right\rangle = 4D_{\mathrm{T}} \tau + 2 V^2 \tau_{\mathrm{R}}^2 \left( \tau / \tau_{\mathrm{R}} + e^{- \tau / \tau_{\mathrm{R}}} - 1 \right).
	\label{eq:ABP-MSD}
\end{equation}
to obtain $V$ = \SI{1.02}{\um/\s} and $\tau_{\mathrm{R}}$ = \SI{2.61}{\second}.

\begin{figure}[hbtp]
	\includegraphics[width=0.5\textwidth]{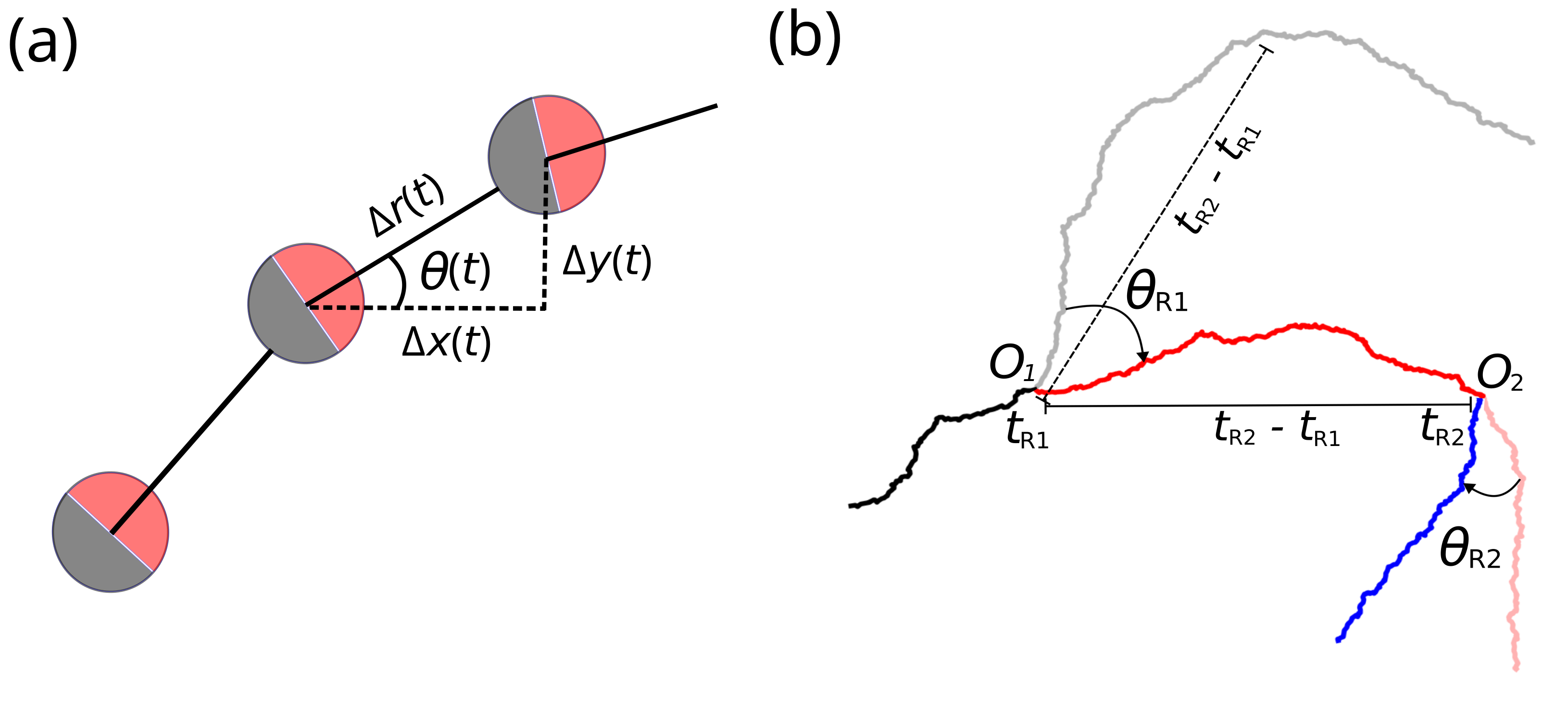}
	\caption{Implementation of SOR by applying rotational transformation. (a) The schematic shows instantaneous orientation $\theta (t)$ and radial displacement $\Delta r (t) = r(t) - r(t - \Delta t)$. (b) Implementation of two SOR events, R1 at $t_{\mathrm{R1}}$ and R2 at $t_{\mathrm{R2}}$, are shown, where the subsequent part of an ABP trajectory (grey, light red) is rotated by reset angles $\theta_{\mathrm{R1}}$ and $\theta_{\mathrm{R2}}$ around $O_1$ and $O_2$, respectively, with the corresponding resultant trajectory segments shown in red and blue.} 
	\label{fig:SORImplement}
\end{figure}

\paragraph{Implementation of SOR.}
We implemented SORs by applying rotational transformations to the subsequent ABP motion after exponentially distributed reset intervals $\Delta t_{\mathrm{reset}}$ with a mean value of $1/\lambda$ (Fig. \ref{fig:SORImplement}). Reset angles $\theta_{\mathrm{reset}}$ were randomly chosen from a uniform distribution bounded by a \textit{reset-cone} of angular width $2\phi$ (Fig. \ref{fig:OR}(b)). The position of the ABP at an SOR instant (\textit{e.g.}, R1 at $t = t_{\mathrm{R1}}$) was set as center $O_1 \equiv \left( x (t_{\mathrm{R1}}), y(t_{\mathrm{R1}})\right)$ to apply an instantaneous orientational transformation by angle $\theta_{\mathrm{R1}} = \theta_{\mathrm{reset}}$ on the later part of the active dynamics, which was updated as $x_{\mathrm{R1}}(t) = x(t) - x(t_{\mathrm{R1}})$, $y_{\mathrm{R1}}(t) = y(t) - y(t_{\mathrm{R1}})$. Thus, we obtained $r (t \geq t_{\mathrm{R1}}) = \sqrt{x_{\mathrm{R1}}^2 (t) + y_{\mathrm{R1}}^2 (t)}$ with respect to $O_1$ and $\theta (t \geq t_{\mathrm{R1}}) = \theta(t) + \theta_{\mathrm{R1}}$, where $\theta(t)$ was computed as $\theta(t) = \tan^{-1}(\Delta y(t) / \Delta x(t))$ (Fig. \ref{fig:SORImplement}(a)). $\Delta x(t) = x (t) – x (t- \Delta t)$, where $\Delta t$ is the time interval between two consecutive frames. By repeating this process, we obtained the updated ABP dynamics after SORs (Fig. \ref{fig:SORImplement}(b)). Finally, we converted the instantaneous displacements $\Delta r (t)$ to Cartesian coordinates as $\Delta x (t) = \Delta r (t) \cos (\theta (t))$ and $\Delta y (t) = \Delta r (t) \sin (\theta (t))$ to obtain the resultant trajectory $\left( x (t), y (t)\right)$ for further analysis and comparison with RnT motions.

\paragraph{Effect of positional fluctuation.}
The instantaneous displacements $\Delta x (t)$ and $\Delta y (t)$, which are used to calculate $\theta (t)$ (Fig. \ref{fig:SORImplement}(a)), comprise both active displacement ($\bm{V} \Delta t$) and Brownian translational fluctuations. Therefore, at a lower value of $V$, $\theta (t)$ no longer reliably represents the intrinsic direction of propulsion of the ABP because of comparatively significant positional fluctuations. This affects the application of SORs and becomes critical in the anisotropic implementation of SOR to replicate taxis. Hence, the MSDs of the resultant dynamics under taxis-mimicking-SOR at $V$ = \SI{1.02}{\um/\s} differ from the corresponding theoretical predictions, whereas match perfectly at $V$ = \SI{12.0}{\um/\s} (Fig. \ref{fig:Chemotaxis}(c, d)).

\begin{figure}[hbtp]
	\includegraphics[width=0.5\textwidth]{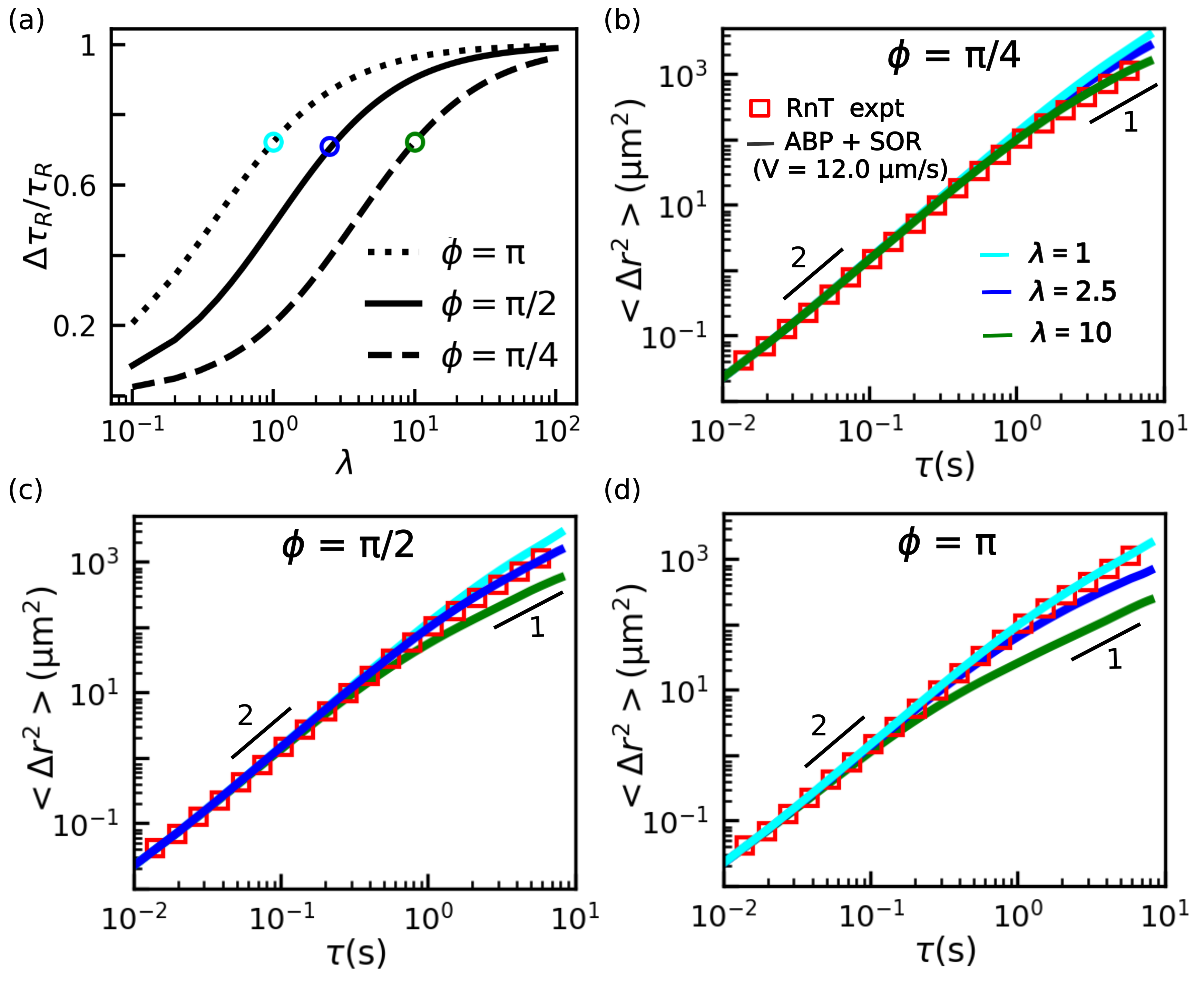}
	\caption{Variation in relative reduction in persistence ($\Delta \tau_{\mathrm{R}} / \tau_{\mathrm{R}}$) with $\phi$ and $\lambda$. (a) $\Delta \tau_{\mathrm{R}} / \tau_{\mathrm{R}}$ is plotted against $\lambda$ for $\phi$ = $\pi$, $\pi$/2, and $\pi$/4, following Eq. \ref{eq: RelPersistence}. (b -d) Resultant MSDs (solid lines) from simulated ABP dynamics with $V$ = \SI{12.0}{\um/\s} under SOR with varied $\phi$ ($\pi$/4, $\pi$/2 and $\pi$) and $\lambda$ (1, 2.5, and 10) are compared with that of RnT dynamics of \textit{E. coli} (red open squares). The $\Delta \tau_{\mathrm{R}} / \tau_{\mathrm{R}}$ values of the resultant MSDs that match the RnT MSD are marked with color-coded circles $\phi$ in (a).}
	\label{fig:VariationPhiLambda}
\end{figure}

\paragraph{Effects of $\phi$ and $\lambda$ in shortening the persistence.}
The dependence of the relative shortening of the persistence $\Delta \tau_{\mathrm{R}} / \tau_{\mathrm{R}}$ on $\phi$ and $\lambda$ is expressed by Eq. \ref{eq: RelPersistence} and are plotted in Fig. \ref{fig:VariationPhiLambda}(a) for three values of $\phi$. While the same $\Delta \tau_{\mathrm{R}} / \tau_{\mathrm{R}}$ can be achieved by choosing suitable sets of $\phi$ and $\lambda$, as marked by the color-coded circles in Fig. \ref{fig:VariationPhiLambda}(a), a larger value of $\phi$ shortens the persistence to a greater extent, with a higher value of $\Delta \tau_{\mathrm{R}} / \tau_{\mathrm{R}}$ at the same $\lambda$. This is because a wider \textit{reset-cone} allows larger values of $\theta_{\mathrm{reset}}$, and hence reduces the persistence more effectively at the same reset rate. All the $\Delta \tau_{\mathrm{R}} / \tau_{\mathrm{R}}$ curves become flatter and approach $\Delta \tau_{\mathrm{R}} / \tau_{\mathrm{R}}$ = 1, \textit{i.e.}, the persistence is completely lost after an adequately high value of $\lambda$, which is smaller for a wider \textit{reset-cone}. These are further demonstrated by comparing the resultant MSDs of ABP with $V$ = \SI{12.0}{\um/\s} under SOR at varying $\phi$ and $\lambda$ with that of the RnT (Fig. \ref{fig:VariationPhiLambda}(b-d)). The required reduction in persistence in the ABP dynamics to match the RnT motion is achieved at $\lambda$ = 10, 2.5, and 1 for $\phi$ = $\pi$/4, $\pi$/2, and $\pi$, respectively, as indicated by the color-coded circles in Fig \ref{fig:VariationPhiLambda}(a). Furthermore, the effective reduction in persistence for the same values of $\lambda$ becomes progressively larger as $\phi$ increases from $\pi$/4 to $\pi$ (Fig. \ref{fig:VariationPhiLambda}(b-d)).

\end{document}


\title{Supplemental Material for \\``Emulating microbial run-and-tumble and tactic motion by stochastically reorienting synthetic active Brownian particles''}


\author{Sandip Kundu}
\affiliation{Department of Physics, Indian Institute of Technology Kanpur, Kanpur - 208016, India}

\author{Dibyendu Mondal}
\affiliation{Department of Physics, Indian Institute of Technology Kanpur, Kanpur - 208016, India}

\author{Arup Biswas }
\affiliation{The Institute of Mathematical Sciences, CIT Campus, Taramani, Chennai - 600113, India}
\affiliation{Homi Bhabha National Institute, Training School Complex, Anushakti Nagar, Mumbai 400094, India}

\author{Arnab Pal}
\email[Contact author: ]{arnabpal@imsc.res.in}
\affiliation{The Institute of Mathematical Sciences, CIT Campus, Taramani, Chennai - 600113, India}
\affiliation{Homi Bhabha National Institute, Training School Complex, Anushakti Nagar, Mumbai 400094, India}

\author{Manas Khan}
\email[Contact author: ]{mkhan@iitk.ac.in}
\affiliation{Department of Physics, Indian Institute of Technology Kanpur, Kanpur - 208016, India}


\maketitle

\onecolumngrid

This Supplemental Material presents the experimental procedures employed in the main text, together with a comprehensive account of the theoretical formalism and derivations. It further includes additional discussions that substantiate and complement the analytical results reported in the Letter.

\tableofcontents
\newpage

\section{Experimental details}

\subsection{Supplementary figures}

\begin{figure*}[h!]
    \centering
    \includegraphics[width=0.7\linewidth]{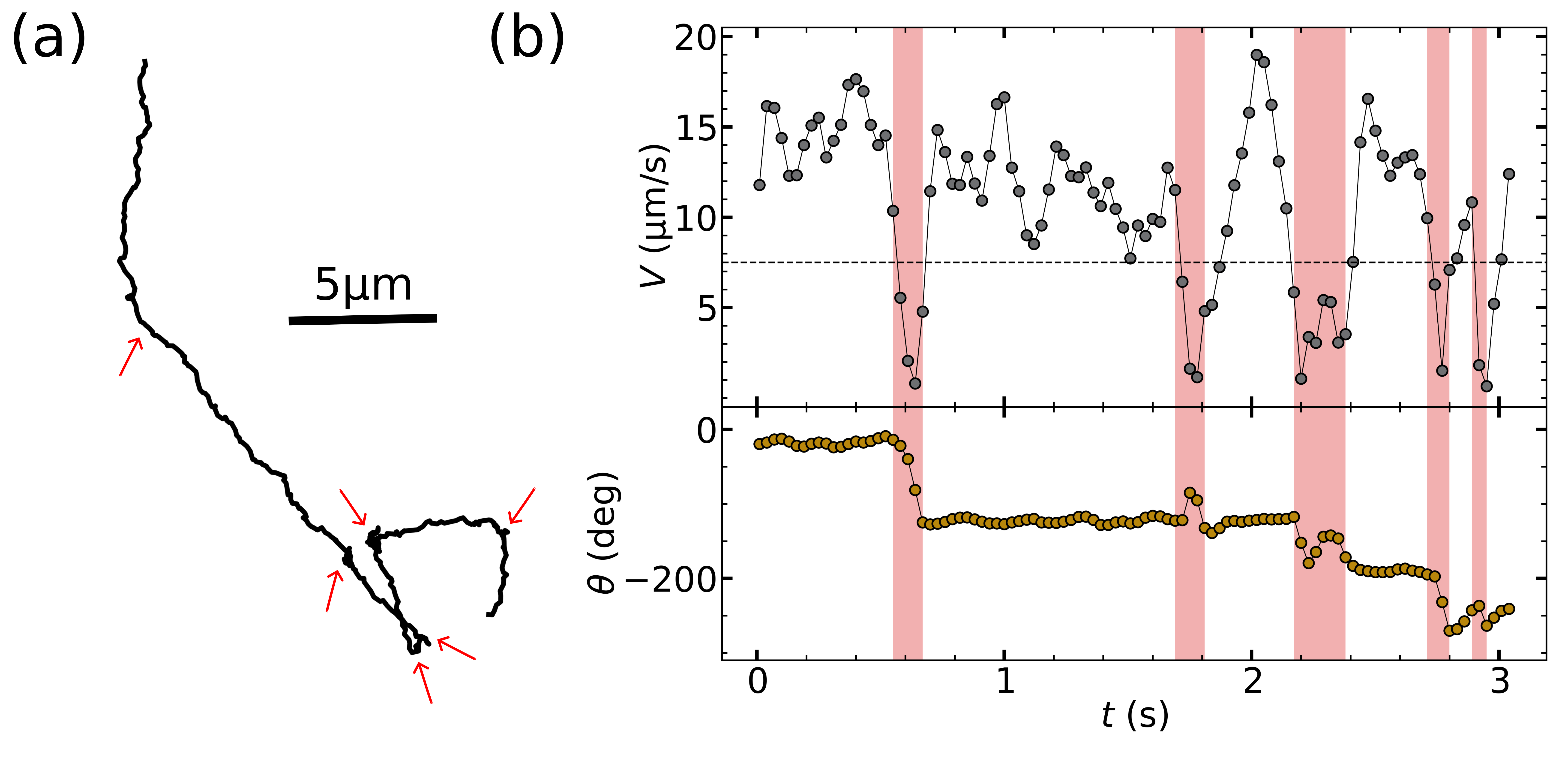}
    \caption{Identification of run and tumble states in RnT trajectory of \textit{E. coli}. (a) A typical RnT trajectory is shown, where sudden angular changes are indicated by red arrows. (b) The instantaneous speed ($V$) and direction ($\theta$), which are obtained from the trajectory, are plotted against time, $t$. The run and tumble states (pink shades) are identified when the variations in both $V$ and $\theta$ satisfy the set criteria (see End Matter for details). A characteristic value of $V$ = \SI{7.5}{\um/\s}, used in the identification criteria, is marked by a dashed horizontal line.} 
		\label{fig:RunTumbleIdentification}
\end{figure*}


\begin{figure*}[h!]
	\centering
	\includegraphics[width=0.62\textwidth]{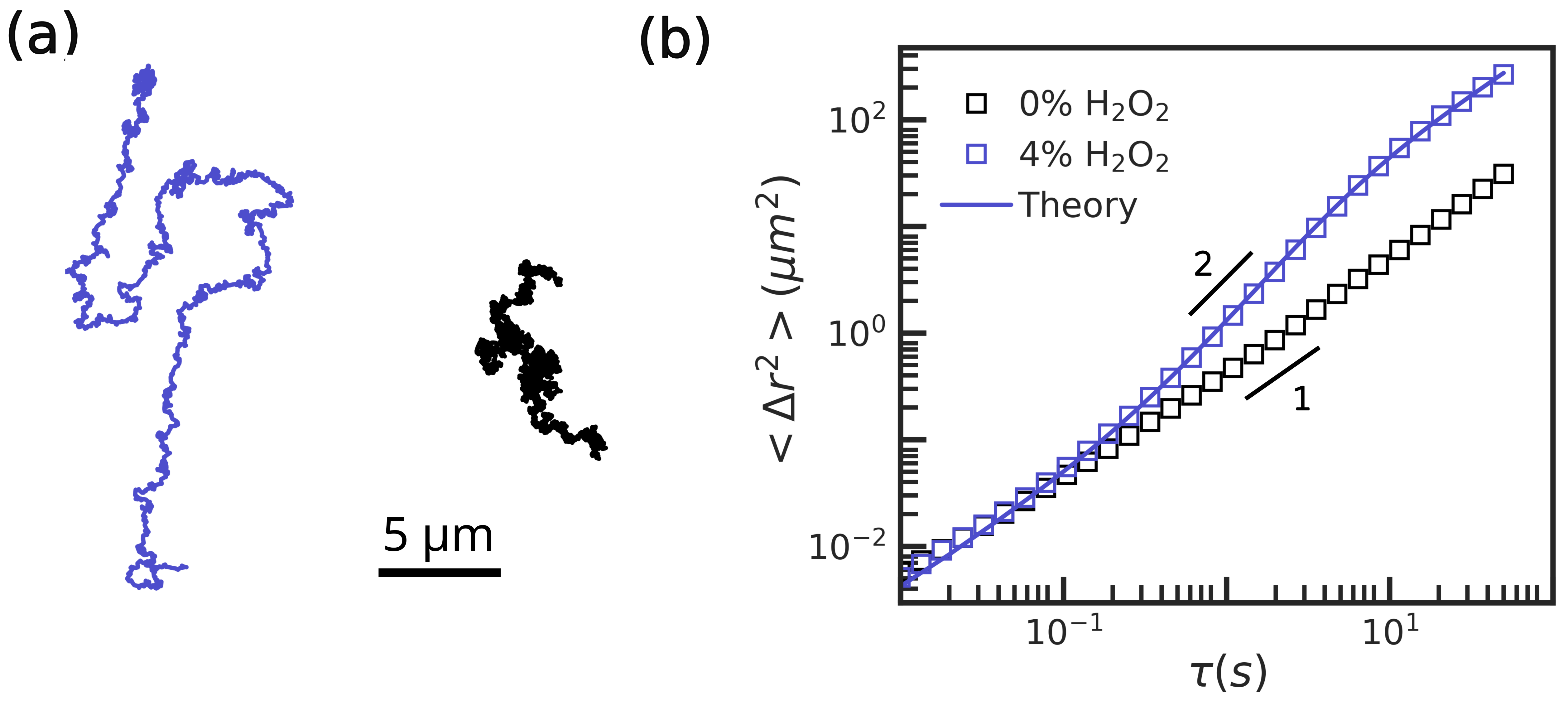}
	\caption{ABP dynamics compared to Brownian diffusion. (a) A typical experimentally captured trajectory of half-Pt-coated silica Janus microparticles (diameter \SI{1.67}{\um}) in 4\% (($v/v$) aqueous suspension of $\mathrm{H}_2\mathrm{O}_2$ (blue) is shown alongside the Brownian diffusion of a similar Janus colloid in absence of $\mathrm{H}_2\mathrm{O}_2$. (b)  Corresponding time- and ensemble-averaged MSDs ($\left\langle \Delta r^2\right\rangle$) are plotted against the time lag $\tau$ with open squares of the same colors as those of the trajectories. The blue curve represents the fit of the ABP MSD (blue open squares) to the analytical prediction (Eq. 5 in the End Matter), which provides the average propulsion speed $V$ and persistence time $\tau_{\mathrm{R}}$. The small straight black lines with apparent slopes of 2 and 1 indicate the corresponding exponents.}
	\label{ABPCharacterization}
\end{figure*}

\subsection{Culture of \textit{E. Coli}}

\textit{E. coli} (wild type, strain RP437) were grown overnight in LB medium at \SI{37}{\celsius} and 175 rpm in an incubator shaker after inoculation from a single colony grown on a LB agar plate. A small volume (\SI{100}{\ul}) of the overnight culture was inoculated with \SI{10}{\ml} of tryptone medium (1\% tryptone and 0.5 \% NaCl) and allowed to grow at 37 \SI{37}{\celsius} and 175 rpm to the mid-logarithmic phase. The cultured bacteria were washed thrice by centrifugation (at 2500 rpm for \SI{5}{\minute}) and then re-suspension in motility buffer (10 mM potassium phosphate buffer, pH 7.5) containing 0.1 mM EDTA, which facilitate sustained motility.

\section{Active Brownian Motion with orientation resets: Theoretical framework and results}\label{sc1}
In this section, we provide the technical details involved in computing the MSD of active Brownian particles under SOR protocols. 

We start by recalling the rudimentaries of the ABP dynamics, in the absence of translational diffusion. Later, we will include translational diffusion to the dynamics (which will be used to corroborate with the experiments) and discuss the changes in the statistical properties. Henceforth, 
the equations of motion of an ABP in two-dimensions, in the absence of translation diffusion, is given by 

\begin{align}
   & \frac{dx}{dt}=V \cos\theta(t),\nonumber\\
     & \frac{dy}{dt}=V \sin\theta(t), \label{abp}\\
     & \frac{d\theta}{dt}= \eta (t) \nonumber,
\end{align}
where $\eta (t)$ is a white noise with zero mean and correlation $\langle \eta (t) \eta(t') \rangle = 2 D_\text{R} \delta(t-t')$. Let us define the angular probability distribution function (PDF) as $P(\theta,t|\theta_0,t_0)$ which is the conditional probability that the ABP will have an orientation $\theta$ at time $t$ given its initial orientation $\theta_0$ at time $t_0$. This is given by the Gaussian distribution 
\begin{align}
    P(\theta,t|\theta_0,t_0)=\frac{1}{\sqrt{4 \pi D_{\text{R}} (t-t_0)}} \exp \left[-\frac{(\theta-\theta_0)^2}{4 D_{\text{R}} (t-t_0)}\right]. \label{gauss}
\end{align}
Evidently, the propagator will be different when subject to SOR protocols, and let us denote that orientational propagator as $P_{\lambda}(\theta,t|\theta_0,t_0)$. In what follows, we derive this propagator for the two distinct SOR protocols and leverage them to obtain the spatial moments. Furthermore, we assume that the resetting events occur with a rate $\lambda$ so that the resetting times are drawn from an exponential distribution given by
\begin{align}
    f_\text{R}(t)=\lambda e^{-\lambda t}. \label{exp}
\end{align}
This implies the following microscopic evolution of the orientation at each time interval $\Delta t$
\begin{eqnarray}
\theta(t+\Delta t)=
\begin{cases}
 \theta(t) + \sqrt{2 D_{\text{R}} \Delta t}~\eta (t)  &\text{with probability }~ 1-r\Delta t, \vspace{0.1cm}\\ 
\theta(t) + \theta_{\mathrm{reset}} &\text{with probability}~ r\Delta t ~(\text{case I}),\\
 \theta_{\mathrm{reset}} &\text{with probability}~ r\Delta t ~(\text{case II}),
\end{cases}
\end{eqnarray}
where $\theta_{\mathrm{reset}}$ is an uniformly distributed random variable inside the interval $[-\phi,\phi]$ and $\eta(t)$ is a Dirac-delta correlated Gaussian noise with zero mean and unit variance. 
Throughout the derivation, we shall assume the ABP starts its motion from the origin \textit{i.e.} $(x,y)\equiv (0,0)$.

\begin{figure}
    \centering
    \includegraphics[width=16cm]{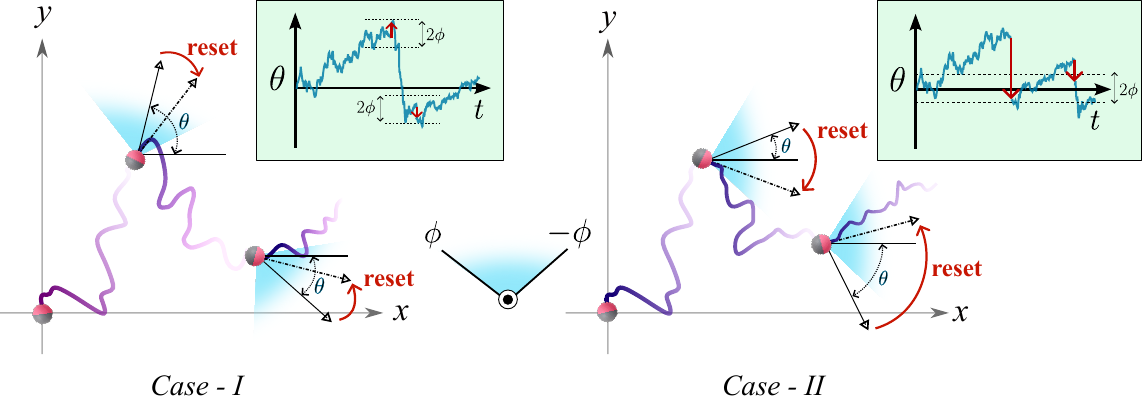}
    \caption{Schematic depiction of two different resetting protocols. Left panel: For case-I, the orientation of the ABP is reset inside the cone $[-\phi,\phi]$ centred around the instantaneous angle of the ABP. A sample stochastic trajectory of the orientational angle $\theta(t)$ under this reset protocol is shown in the inset. Right panel: For case II, the cone is centred parallel to the $x$-axis. The inset shows a typical trajectory of the $\theta(t)$ for this particular reset protocol.}
    \label{fig1}
\end{figure}

\subsection{Case I: Reset to uniform distribution with respect to instantaneous orientation: Emulating RnT dynamics}
As mentioned in the main text we apply the SOR protocols to the ABP trajectories by reorienting the propulsion to a new direction which is chosen randomly from a uniform distribution bounded by a \textit{reset-cone} of angular width $2\phi$, after exponentially distributed time intervals with reset-rate $\lambda$. In here, we first assume that reset occurs uniformly within the reset cone. In other words, if $\tau_f$ denotes the time epoch of resetting then the angle of orientation of the ABP after the resetting event is given by a random angle in the interval $[\theta(\tau_f)-\phi,\theta(\tau_f)+\phi]$. In this case the renewal equation for the propagator $P_{\lambda}(\theta,t|\theta_0,t_0)$ can be written, following the first-renewal formalism  \cite{pal2016diffusion}, as
\begin{align}
    P_{\lambda}(\theta,t|\theta_0,t_0)=e^{-\lambda t} P(\theta,t|\theta_0,t_0) + \lambda\int_{t_0}^t d\tau_f  e^{-\lambda \tau_f} P_{\lambda}(\theta,t|\Theta,\tau_f), \label{ren-2}
\end{align}
where $\Theta \in [\theta(\tau_f)-\phi,\theta(\tau_f)+\phi]$ is introduced to indicate the random angle uniformly chosen from the interval. Several comments are in order to elaborate the structure of the renewal equation. The first term on the RHS of this equation accounts for the events for which no resetting has occurred upto time $t$. Since the resetting times are drawn from an exponential distribution as in \eref{exp}, the probability of having no resetting upto time $t$ is simply $e^{-\lambda t}$. Thus, the contribution to the overall propagator simply comes from the reset-free process's propagator which is $P(\theta,t|\theta_0,t_0)$ multiplied with the exponential factor mentioned in above. On the other hand, the second term in the RHS of the renewal equation \eref{ren-2} is the contribution emanating from those trajectories that undergo atleast one orientational resetting. Following a first resetting event at time $\tau_f$ with probability $\lambda d\tau_f$, the orientation angle restarts from a new angle $\Theta$ given by
\begin{align}
    \Theta =\theta(\tau_f) + \theta_{\mathrm{reset}},
\end{align}
where $\theta_{\mathrm{reset}}$ is a uniformly distributed random variable in the interval $\in [-\phi,+\phi]$ and the angle $\theta(\tau_f)$ is the instantaneous angle of the ABP at time $\tau_f$. Since the orientational dynamics restarts, we multiply with the propagator $P_{\lambda}(\theta,t|\Theta,\tau_f)$ thus completing the renewal structure.
The second term on the RHS of the renewal equation \eref{ren-2} can be written as
\begin{align}
   \int_{t_0}^t d\tau  e^{-\lambda \tau} P_{\lambda}(\theta,t|\Theta,\tau)&=  \int_{t_0}^t d\tau  e^{-\lambda \tau} P_{\lambda}(\theta,t|\theta(\tau)+\theta_{\mathrm{reset}},\tau) \nonumber\\
   &=\int_{t_0}^t d\tau  e^{-\lambda \tau} \int_{-\infty}^{\infty} d\theta'  P(\theta',\tau|\theta_0,t_0)~ P_{\lambda}(\theta,t|\theta'+\theta_{\mathrm{reset}},\tau),
\end{align}
where we have simply used $\tau$ instead of $\tau_f$ since it is a dummy variable of the equation. Note that in this step we have averaged out the angle $\theta(\tau)$ which occurs with probability $P(\theta',\tau|\theta_0,t_0)$ given by \eref{gauss}. From here it is clear that $ P_{\lambda}(\theta,t|\theta_0,t_0)$ is homogeneous when $ P(\theta,t|\theta_0,t_0)$ is also homogeneous. That implies one can write $  P_{\lambda}(\theta,t|\theta'+\theta_{\mathrm{reset}},\tau)=P_{\lambda}(\theta-\theta' -\theta_{\mathrm{reset}},t-\tau|0,0)$. For convenience and without any loss of generality from now on we shall set $t_0=0$ and denote $P_{\lambda}(\theta,t|\theta_0,0)= P_{\lambda}(\theta - \theta_0,t|0,0) \equiv P_{\lambda}(\theta -\theta_0,t)$. With this identification we have
\begin{align}
     \int_0^t d\tau  e^{-\lambda \tau} P_{\lambda}(\theta,t|\Theta,\tau)&=\int_0^t d\tau  e^{-\lambda \tau} \int_{-\infty}^{\infty} d\theta'  P(\theta'-\theta_0,\tau)~ P_{\lambda}(\theta-\theta' -\theta_{\mathrm{reset}},t-\tau).
\end{align}
Next, we average out the angle $\theta_{\mathrm{reset}}$ with respect to a uniform distribution to have 
\begin{align}
     \int_0^t d\tau  e^{-\lambda \tau} P_{\lambda}(\theta,t|\Theta,\tau)&=\int_0^t d\tau  e^{-\lambda \tau} \int_{-\infty}^{\infty} d\theta'  P(\theta'-\theta_0,\tau)~ \int_{-\phi}^{\phi} d\phi' ~ \frac{1}{2\phi} P_{\lambda}(\theta-\theta' -\phi',t-\tau).
\end{align}
The complete renewal equation \eref{ren-2} now takes the following form 
\begin{align}
    P_{\lambda}(\theta-\theta_0,t)&=e^{-\lambda t} P(\theta -\theta_0,t) + \lambda \int_0^t d\tau  e^{-\lambda \tau} \int_{-\infty}^{\infty} d\theta'  P(\theta'-\theta_0,\tau)~ \int_{-\phi}^{\phi} d\phi' ~ \frac{1}{2\phi} P_{\lambda}(\theta-\theta' -\phi',t-\tau) \nonumber\\
    &=e^{-\lambda t} P(\theta-\theta_0,t) + \frac{\lambda}{2\phi} \int_{-\phi}^{\phi} d\phi' \int_0^t d\tau  e^{-\lambda \tau} \int_{-\infty}^{\infty} d\theta'  P(\theta'-\theta_0,\tau)~   P_{\lambda}(\theta-\theta' -\phi',t-\tau).
    \label{ren-22}
\end{align}
For making further progress, it will prove convenient to define $\theta(t)-\theta_0=\overline{\theta}$ as the displacement angle so that 
\begin{align}
     P_{\lambda}(\overline{\theta},t) =e^{-\lambda t} P(\overline{\theta},t) + \frac{\lambda}{2\phi} \int_{-\phi}^{\phi} d\phi' \int_0^t d\tau  e^{-\lambda \tau} \int_{-\infty}^{\infty} d\theta'  P(\theta',\tau)~   P_{\lambda}(\overline{\theta}-\theta' -\phi',t-\tau)
\end{align}

Let us now take the Fourier transform of the above equation defined by $\widetilde{P}_{\lambda}(k,t)= \int_{-\infty}^{\infty} d\overline{\theta} ~e^{i k \overline{\theta}}P_{\lambda}(\overline{\theta},t) $ to have
\begin{align}
    \widetilde{P}_{\lambda}(k,t)=e^{-\lambda t}  \widetilde{P}(k,t) + \frac{\lambda}{2\phi} \int_{-\phi}^{\phi} d\phi' \int_0^t d\tau  e^{-\lambda \tau}  e^{i k \phi'} \widetilde{P}(k,\tau)~ \widetilde{P}_{\lambda}(k,t-\tau).
\end{align}
Here we have used the convolution property of the Fourier transform in the second term. Furthermore, taking the Laplace transform of the above equation yields
\begin{align}
     &\widetilde{P}_{\lambda}(k,s)=  \widetilde{P}(k,\lambda+s) + \frac{\lambda}{2\phi} \int_{-\phi}^{\phi} d\phi' ~ e^{i k \phi'}  \widetilde{P}(k,\lambda+s)~ \widetilde{P}_{\lambda}(k,s) 
     \end{align}
After some rearrangement, we obtain
     \begin{align}
     \widetilde{P}_{\lambda}(k,s)=\frac{ \widetilde{P}(k,\lambda+s) }{1- \lambda \widetilde{P}(k,\lambda+s) \left(\frac{\sin k \phi}{k\phi} \right)}. \label{fl}
\end{align}
Note that the propagator \eref{gauss} in the combined Fourier-Laplace space can be found to be
\begin{align}
     \widetilde{P}(k,s)=\frac{1}{D_{\text{R}} k^2 +s}.
\end{align}
Substituting this result in \eref{fl} we find the reset induced propagator in Fourier-Laplace space given by
\begin{align}
    \widetilde{P}_{\lambda}(k,s)=\frac{1}{D_{\text{R}} k^2 +\lambda +s - \lambda\left(\frac{\sin k \phi}{k\phi} \right) }.
\end{align}
The above result can be easily Laplace inverted to yield the propagator in the Fourier space as
\begin{align}
    \widetilde{P}_{\lambda}(k,t)=e^{-\left[D_{\text{R}} k^2 +\lambda- \lambda\left(\frac{\sin k \phi}{k\phi} \right)  \right]t }. \label{prp-f}
\end{align}
The propagator in the Fourier space as found in \eref{prp-f} will be sufficient to evaluate the first two moments of the position of the ABP as we show below.
\subsubsection{\textbf{Mean position}}
The exact expression for the mean found from the set of equations \eref{abp} is given by
\begin{align}
   & \langle x (t) \rangle = V \int_0^t dt' \langle \cos \theta(t') \rangle, \label{mx}\\
     & \langle y (t) \rangle = V \int_0^t dt' \langle \sin \theta(t') \rangle \label{my}.
\end{align}
To find the angular expectations inside the integrals we observe that in the limit $k\to 1$ the propagator can be written as
\begin{align}
     \widetilde{P}_{\lambda}(k\to 1,t)&= \int_{-\infty}^{\infty} d\theta ~e^{i  \theta}P_{\lambda}(\theta,t|\theta_0) \nonumber\\
      &=e^{i  \theta_0}\int_{-\infty}^{\infty} d(\theta-\theta_0) ~e^{i  (\theta-\theta_0)}P_{\lambda}(\theta-\theta_0,t) \nonumber\\
      &=e^{i  \theta_0}\int_{-\infty}^{\infty} d\theta ~e^{i  \theta}P_{\lambda}(\theta,t) \nonumber\\
     &=(\cos \theta_0 + i \sin \theta_0)\int_{-\infty}^{\infty} d\theta ~(\cos \theta + i \sin \theta)P_{\lambda}(\theta,t) \nonumber\\
     &=\left(A\cos \theta_0-B \sin \theta_0\right)+i (A \sin \theta_0 +B \cos \theta_0),
\end{align}
where $A=\int_{-\infty}^{\infty} d\theta ~\cos \theta P_{\lambda}(\theta,t)$ and $B=\int_{-\infty}^{\infty} d\theta ~\sin \theta P_{\lambda}(\theta,t)$.
Now setting $k=1$ in \eref{prp-f} we find
\begin{align}
    A+i B=e^{-\left[D_{\text{R}} +\lambda+ \lambda\left(\frac{\sin \phi}{\phi} \right)  \right]t }.
\end{align}
This in turn gives
\begin{align}
     &A=e^{-\left[D_{\text{R}} +\lambda- \lambda\left(\frac{\sin \phi}{\phi} \right)  \right]t }, \\
     &B=0.
\end{align}
Combined we have
\begin{align}
    &\langle \cos \theta(t) \rangle=\left(A\cos \theta_0-B \sin \theta_0\right)=e^{-\left[D_{\text{R}} +\lambda+ \lambda\left(\frac{\sin \phi}{\phi} \right)  \right]t } \cos\theta_0 \\
    & \langle \sin \theta(t) \rangle=\left(A\cos \theta_0-B \sin \theta_0\right)=e^{-\left[D_{\text{R}} +\lambda+ \lambda\left(\frac{\sin \phi}{\phi} \right)  \right]t } \sin\theta_0
\end{align}
Plugging these expressions back to \eref{mx}-(\ref{my}) we finally obtain the expression for the mean of the position as 
\begin{align}
    &\langle x(t) \rangle_{\theta_0}=\left[\frac{V \left(1 -  e^{-\left[D_{\text{R}} +\lambda- \lambda\left(\frac{\sin \phi}{\phi} \right)  \right]t }\right)}{ D_{\text{R}}+\lambda- \lambda \left(\frac{\sin \phi}{\phi}\right)}\right]\cos\theta_0 \label{1stmom},\\
    &\langle y(t) \rangle_{\theta_0}=\left[\frac{V \left(1 -  e^{-\left[D_{\text{R}} +\lambda- \lambda\left(\frac{\sin \phi}{\phi} \right)  \right]t }\right)}{ D_{\text{R}}+\lambda- \lambda \left(\frac{\sin \phi}{\phi}\right)}\right]\sin\theta_0. 
\end{align}
Note that one can define a resetting rate dependent time-scale $\tau_{\lambda}=\left( D_{\mathrm{R}}+\lambda-\lambda \frac{\sin(\phi)}{\phi} \right)^{-1}$ in terms of which the above quantities takes a simpler form as given below
\begin{align}
    &\langle x(t) \rangle_{\theta_0}=V \tau_\lambda (1- e^{-t/\tau_\lambda })\cos\theta_0, \\
    &\langle y(t) \rangle_{\theta_0}=V \tau_\lambda (1- e^{-t/\tau_\lambda})\sin\theta_0, 
\end{align}
where we have introduced a resetting dependent persistence time 
\begin{align}
    \tau_\lambda = \left( D_{\mathrm{R}}+\lambda-\lambda \frac{\sin(\phi)}{\phi} \right)^{-1},
\end{align}
which was also used in the main text.

\subsubsection{\textbf{Variance of the position}}
Let us now proceed to find the variance of the position of the ABP. Let us focus first on the second moment of $x$. Following \eref{abp}, this can be written as
\begin{align}
    \langle x^2 (t) \rangle= V^2 \int_0^t dt' \int_0^t dt''\langle \cos \theta(t') \cos \theta(t'')\rangle. \label{2m}
\end{align}
Evidently, to find the second moment we need to compute the angular correlations. For convenience of notations let us denote $\theta(t')=\theta'$ and $\theta(t'')=\theta''$. We then need to compute the correlation $\langle \cos \theta' \cos \theta'' \rangle$ which can also be written as
\begin{align}
    \langle \cos \theta' \cos \theta'' \rangle&=\frac{1}{4}\left\langle\left(e^{i \theta'}+e^{-i \theta'}\right)\left(e^{i \theta''}+e^{-i \theta''}\right)\right\rangle \nonumber\\
    &=\frac{1}{4} \left(\left\langle e^{i (\theta'+\theta'')}\right\rangle +\left\langle e^{i (\theta'-\theta'')}\right\rangle+\left\langle e^{-i (\theta'-\theta'')}\right\rangle+\left\langle e^{-i (\theta'+\theta'')}\right\rangle\right). \label{eit}
\end{align}
Let us try to evaluate one of the expectations under the parentheses of the above equation. Assuming $t''>t'$ we find
\begin{align}
   \left\langle e^{i (k'\theta'+k''\theta'')}\right\rangle &=\int_{-\infty}^{\infty} d\theta'' ~e^{ik''\theta''} \int_{-\infty}^{\infty} d\theta'~  e^{ik'\theta'}  P_{\lambda}(\theta'',t''|\theta',t') P_{\lambda}(\theta',t'|\theta_0) ~~ \text{for }~~t''>t' \nonumber\\
   &=\int_{-\infty}^{\infty} d\theta'' ~e^{ik''\theta''} \int_{-\infty}^{\infty} d\theta'~  e^{ik'\theta'}  P_{\lambda}(\theta''-\theta',t''-t') P_{\lambda}(\theta'-\theta_0,t'). 
   \end{align}
One can again use the convolution property of Fourier transform in the above equation to obtain
\begin{align}
    \left\langle e^{i (k'\theta'+k''\theta'')}\right\rangle&=  \widetilde{P}_{\lambda}(k'',t''-t') \widetilde{P}_{\lambda}(k'+k'',t') e^{2i (k'+k'') \theta_0}.
\end{align}
By setting $k'=k''=1$ and using \eref{prp-f} one then obtains the exact expression for the correlation as
\begin{align}
     \left\langle e^{i (\theta'+\theta'')}\right\rangle&=  e^{2i \theta_0}\widetilde{P}_{\lambda}(1,t''-t') \widetilde{P}_{\lambda}(2,t')=e^{2i \theta_0}e^{-\left[D  + \lambda- \lambda\left(\frac{\sin \phi}{\phi} \right)  \right][t''-t'] }\times e^{-\left[4D_{\text{R}} +\lambda- \lambda\left(\frac{\sin 2 \phi}{2\phi} \right)  \right]t' }.
\end{align}
Using similar approach one can obtain all the other three correlations in \eref{eit} and finally arrives at the following result for the correlation 
\begin{align}
    \langle \cos \theta' \cos \theta'' \rangle =\frac{1}{2} e^{ -t'' (D_{\text{R}}+\lambda)-3 D_{\text{R}} t'+\frac{\lambda (t''-t') \sin \phi}{\phi }}\left(e^{t' (4 D_{\text{R}}+\lambda)}+ e^{\frac{\lambda t' \sin \phi \cos \phi}{\phi }}\cos 2 \theta_0 \right) ~~ \text{for }~~t''>t'. \label{corr}
\end{align}
By reversing the notation $t'$ and $t''$ one obtains the corresponding correlation function for $t''<t'$ as well. Now breaking up \eref{2m} in distinct regions based on $t'$ and $t''$ we find 
\begin{align}
   \langle x^2 (t) \rangle= V^2 \int_0^t dt' \left( \int_0^{t'} dt''\langle \cos \theta(t'') \cos \theta(t')\rangle +\int_{t'}^{t} dt''\langle \cos \theta(t') \cos \theta(t'')\rangle \right). \label{2m-2}
\end{align}
Plugging the result for the correlation as in \eref{corr} we find the exact expression for the second moment of $x$. The exact expression for the second moment is found to be
\begin{align}
\langle x^2 (t) \rangle_{\theta_0}=& -\frac{V^2\phi^2 \cos 2\theta_0 e^{-t (4 D_{\text{R}}+\lambda) } \left(e^{3 D_{\text{R}} t+\frac{\lambda t \sin \phi}{\phi }}-e^{\frac{\lambda t \sin \phi \cos \phi}{\phi }}\right)}
{2 (\phi  (D_{\text{R}}+\lambda)-\lambda \sin \phi) (3 D_{\text{R}}\phi +\sin \phi (\lambda-\lambda \cos \phi))} \nonumber \\
& +\frac{V^2\phi^2 \cos 2 \theta_0  \left(e^{t \left(-4 D_{\text{R}}+\frac{\lambda \sin \phi \cos \phi}{\phi }- \lambda\right)}-1\right)}
{2 (\phi  (4 D_{\text{R}}+\lambda)-\lambda \sin \phi \cos \phi) (3 D_{\text{R}} \phi +\sin \phi (\lambda-\lambda \cos \phi))} \nonumber \\
& -\frac{ V^2\phi^2 \cos 2\theta_0 \left(e^{t \left(-4 D_{\text{R}}+\frac{\lambda \sin \phi \cos \phi}{\phi }- \lambda\right)}-1\right)}
{2 (\phi  (D_{\text{R}}+\lambda)-\lambda \sin \phi) (\phi  (4 D_{\text{R}}+\lambda)-\lambda \sin \phi \cos \phi)}  \nonumber \\
& +\frac{V^2\phi^2  \cos 2 \theta_0  \left(1 -  e^{\frac{\lambda t \sin \phi}{\phi }-t (D_{\text{R}}+\lambda)}\right)}
{2 (\phi  (D_{\text{R}}+\lambda)-\lambda \sin \phi) (3 D_{\text{R}} \phi +\sin \phi (\lambda-\lambda \cos \phi))} \nonumber \\
& +\frac{V^2t \phi }{\phi  (D_{\text{R}}+\lambda)-\lambda \sin \phi} 
- \frac{V^2\phi  \left(\phi -\phi  e^{\frac{\lambda t \sin \phi}{\phi }-t (D_{\text{R}}+\lambda)}\right)}
{2 (\phi  (D_{\text{R}}+\lambda)-\lambda \sin \phi)^2} -\frac{V^2\phi  \left(\phi -\phi  e^{-t \left(D_{\text{R}} -\frac{\lambda \sin \phi}{\phi }+ \lambda\right)}\right)}
{2 (\phi  (D_{\text{R}}+\lambda)-\lambda \sin \phi)^2}. \label{2ndmom}
\end{align}
Finally one can find the variance $\langle \Delta x^2 (t) \rangle_{\theta_0}=\langle x^2(t) \rangle_{\theta_0} -\langle x(t) \rangle_{\theta_0}^2$ with \eref{1stmom} and \eref{2ndmom}. Using a similar approach one can find the moments in the $y$ direction as well.

\subsection{
\textit{Average over initial $\theta_0$}} For experimental conditions, the initial orientation $\theta_0$ is chosen uniformly from the interval $[0,2\pi]$.  For this purpose one needs to take the expectation of the mean and second moment over this range of $\theta_0$. After performing this expectation we find 
\begin{align}
    &\langle x(t) \rangle=\frac{1}{2\pi}\int_0^{2\pi}d\theta_0 ~\langle x(t) \rangle_{\theta_0}=0\\
   &\langle x^2(t) \rangle= \frac{1}{2\pi}\int_0^{2\pi}d\theta_0 ~\langle x^2(t) \rangle_{\theta_0}=\frac{V^2\phi  \left( \phi  (t (D_{\text{R}}+\lambda)-1)-\lambda t \sin \phi+\phi  e^{-(D_{\text{R}}+\lambda)t +\frac{\lambda t \sin \phi}{\phi }}\right)}{(\phi  (D_{\text{R}}+\lambda)-\lambda \sin \phi)^2}. \label{2ndmom-avg}
\end{align}
In this case, as the mean is exactly zero thus the variance $\langle \Delta x^2(t) \rangle$ is same as the second moment \textit{i.e.} $\langle \Delta x^2(t) \rangle=\langle x^2(t) \rangle$. Moreover, as the initial condition is symmetrically chosen from all the directions all the moments of the particle in either $x$ or $y$ direction become exactly the same. Evidently, we can infer that the mean and second moment of the particle's position in the $y$ direction is exactly the same as in the $x$-direction so that $\langle y(t) \rangle=\langle x(t) \rangle=0$ and $\langle y^2(t) \rangle=\langle x^2 (t) \rangle$ as in \eref{2ndmom-avg}. If $\Vec{r}(t)=x(t)\hat{i}+y(t)\hat{j}$ denotes the distance of the particle from the origin then the mean and variance of this quantity is given by
\begin{align}
   & \langle \Vec{r}(t) \rangle=\langle x(t) \rangle \hat{i}+\langle y(t) \rangle \hat{j}=0 \\
   & \langle \Delta r^2(t) \rangle=\langle \Vec{r}(t)\cdot \Vec{r} (t)\rangle- \langle\Vec{r}(t)\rangle\cdot \langle\Vec{r}(t)\rangle=\langle \Delta x^2(t) \rangle +\langle \Delta y^2 (t) \rangle \nonumber\\
   &\hspace{5.28cm}=\frac{2V^2\phi  \left( \phi  (t (D_{\text{R}}+\lambda)-1)-\lambda t \sin \phi+\phi  e^{-(D_{\text{R}}+\lambda)t +\frac{\lambda t \sin \phi}{\phi }}\right)}{(\phi  (D_{\text{R}}+\lambda)-\lambda \sin \phi)^2}.
\end{align}

\subsection{
\textit{Incorporating translational diffusion}} Let us now assume that in addition to the rotational diffusion, the ABPs can also undergo translational diffusion in both the $x$ and $y$ direction with diffusion constant $D_{\text{T}}$. In that case, the modified equation of motion is given by 
\begin{align}
   & \frac{dx}{dt}=V \cos\theta(t) + \xi_x(t),\nonumber\\
     & \frac{dy}{dt}=V \sin\theta(t) +\xi_y(t), \label{abp-trans}\\
     & \frac{d\theta}{dt}= \eta  (t) \nonumber,
\end{align}
where $\xi_x(t)$  (or $\xi_y(t)$) is another white noise with correlation $\langle \xi_{x}(t) \xi_{x}(t') \rangle=2D_{\text{T}} \delta(t-t')$. To find the variance in this case, one can use simple physical reasoning which is: the translational diffusion is completely independent and decoupled from the ABM of the particle. Thus the moments arising from the translational diffusion will simply add up to the earlier results. For instance, the variance due to the translational diffusion in the $x$ (or $y$) direction is $2D_{\text{T}} t$. Thus the net variance of the ABP with translational diffusion is given by
\begin{align}
   & \langle \Delta x^2 (t) \rangle= \langle \Delta y^2 (t) \rangle=\frac{V^2\phi  \left( \phi  (t (D_{\text{R}}+\lambda)-1)-\lambda t \sin \phi+\phi  e^{-(D_{\text{R}}+\lambda)t +\frac{\lambda t \sin \phi}{\phi }}\right)}{(\phi  (D_{\text{R}}+\lambda)-\lambda \sin \phi)^2} + 2D_{\text{T}} t, \\
    &\langle \Delta r^2(t) \rangle=\frac{2V^2\phi  \left( \phi  (t (D_{\text{R}}+\lambda)-1)-\lambda t \sin \phi+\phi  e^{-(D_{\text{R}}+\lambda)t +\frac{\lambda t \sin \phi}{\phi }}\right)}{(\phi  (D_{\text{R}}+\lambda)-\lambda \sin \phi)^2} + 4D_{\text{T}} t, \label{msd-I}
\end{align}
which is our main result. We verify this result against numerical simulations to find an excellent agreement. The results are shown in Fig. \ref{fig1}.

\begin{figure}
    \centering
    \includegraphics[width=0.5\linewidth]{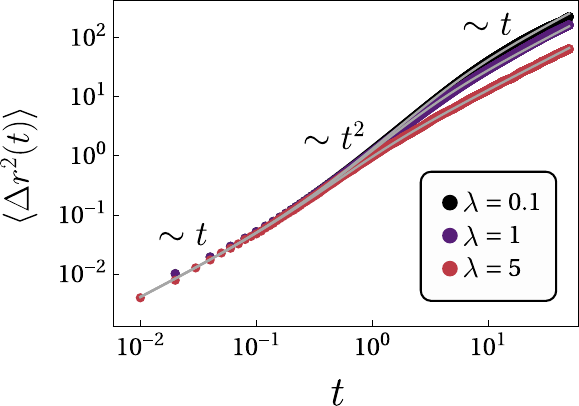}
    \caption{\textbf{Theoretical vs experimental MSD for the case I} with various resetting rate: The circles represent the experimental results and the gray line represents the corresponding result obtained from theory as in \eref{msd-I}. Parameters are: $\phi=\frac{\pi}{2},D_{\text{R}}=\frac{1}{2.61},D_{\text{T}}=0.1,V=1.02$.}
    \label{fig1}
\end{figure}

 From a close inspection of \eref{msd-I} we can again identify the emergence of the resetting rate dependent persistent time scale defined already as $\tau_\lambda= \left[(D_{\text{R}}+\lambda)- \lambda\left(\frac{\sin \phi}{\phi}\right)\right]^{-1}$. The MSD as in \eref{msd-I} can be written in terms of this new timescale $\tau_\lambda$ in a compact form as
\begin{align}
   \langle \Delta r^2(t) \rangle=
   2V^2\tau_\lambda^2 \left( t/\tau_\lambda+e^{-t/\tau_\lambda}-1 \right)+4D_{\text{T}}t,
\label{eq:MSD-th}
\end{align}
which is Eq. (3) in the main text. 
Various asymptotic forms can be derived from the above expression. For instance, at very short times $t\ll \tau_\lambda$, the MSD is given by 
\begin{align}
   \langle \Delta r^2(t \ll \tau_\lambda) \rangle \sim 4D_{\text{T}} t+ V^2 t^2.
\end{align}
Moreover, from here one can identify another time-scale independent of the resetting rate as $t_{0}=\frac{D_{\text{T}}}{V^2}$. Depending on $t_0$ the MSD shows two distinct behaviour 
\begin{eqnarray}
 \langle \Delta r^2(t \ll \tau_\lambda) \rangle\approx
\begin{cases}
 4 D_{\text{T}} t &\text{when }~ t\ll t_{0}, \vspace{0.1cm}\\ 
V^2 t^2  &\text{when }~ t \gg t_0.
\end{cases}
\end{eqnarray}
On the other hand, at very large enough times $t\gg \tau_\lambda$ the MSD has the asymptotic form
\begin{align}
   \langle \Delta r^2(t \gg \tau_\lambda) \rangle \sim 4D_{\text{eff}}~ t,
\end{align}
with $D_{\text{eff}}=D_{\text{T}} + V^2 \tau_\lambda/2 
$ being the effective diffusion coefficient. In summary, we have the complete asymptotic behaviour of the MSD in case-I given by
\begin{eqnarray}
\langle \Delta r^2(t) \rangle\approx
\begin{cases}
 4 D_{\text{T}} t &\text{when }~ t\ll \frac{D_{\text{T}}}{V^2}, \vspace{0.1cm}\\ 
V^2 t^2  &\text{when }~ \frac{D_{\text{T}}}{V^2} \ll t \ll \tau_\lambda 
,\\
4D_{\text{eff}}~ t  &\text{when }~ t \gg \tau_\lambda
.
\end{cases}\label{asymp-I}
\end{eqnarray}
We verified these limits in Fig. \ref{fig1} to find an excellent agreement.

\subsection{Case II: Reset to uniform distribution with respect to origin: Emulating Taxis motion}
To emulate the taxis motion, we now consider a SOR protocol where after each resetting event the angular orientation of the ABP is reset uniformly within the interval $[-\phi,\phi]$ \textit{around the origin}. This implies that the reset-cone is centred around the origin in this case. We proceed using the last-renewal formalism developed in \cite{pal2016diffusion,kumar2020active} where we first assume that the last resetting occurs at time $t-\tau$, and thus, one can write the last renewal equation for the propagator $ P_{\lambda}(\theta,t|\theta_0,t_0)$ as
\begin{align}
    P_{\lambda}(\theta,t|\theta_0,t_0)=e^{-\lambda t} P(\theta,t|\theta_0,t_0) + \lambda \int_{t_0}^t d\tau e^{-\lambda \tau} P(\theta,t|\theta_{\mathrm{reset}},t-\tau). \label{ren-1}
\end{align}
The quantity $\theta_{\mathrm{reset}}$ is a random variable chosen from the uniform distribution between $[-\phi,\phi]$ and $P(\theta,t|\theta_0,t_0)$ is the reset free propagator as in \eref{gauss}.  Let us take the expectation of the variable $\theta_{\mathrm{reset}}$ in \eref{ren-1} to have 
\begin{align}
     P_{\lambda}(\theta,t|\theta_0,t_0)=e^{-\lambda t} P(\theta,t|\theta_0,t_0) + \frac{\lambda}{2\phi} \int_{-\phi}^{\phi} d\phi'\int_{t_0}^t d\tau e^{-\lambda \tau} P(\theta,t|\phi',t-\tau). 
\end{align}
For convenience, we shall set $t_0=0$ from now on. Using the stationary and homogeneous property of the reset-free propagator \textit{i.e.} $P(\theta,t|\phi',t-\tau)=P(\theta-\phi',\tau|0,0)$, one can infer that the resetting induced propagator will also be a stationary one. However, note that the propagator $ P_{\lambda}(\theta,t|\theta_0,t_0)$ is not homogeneous since one can not write a similar renewal equation for $\theta-\theta_0$ as we had done in the previous case. Taking the Fourier transform in the both sides of the above equation we find 
\begin{align}
    \widetilde{P}_{\lambda}(k,t|\theta_0)&=e^{-\lambda t} e^{ik\theta_0}\widetilde{P}(k,t) +\frac{\lambda}{2\phi} \int_{-\phi}^{\phi} d\phi' e^{ik\phi'}\int_{0}^t d\tau e^{-\lambda \tau} \widetilde{P}(k,\tau) \nonumber\\
    &=e^{-\lambda t} e^{ik\theta_0}\widetilde{P}(k,t) + \lambda\left(\frac{\sin k\phi}{k\phi}\right)\int_{0}^t d\tau e^{-\lambda \tau} \widetilde{P}(k,\tau). \label{ren-3}
\end{align}
Here the quantity $\widetilde{P}(k,\tau)$ is the Fourier transform of the propagator as in \eref{gauss} given by 
\begin{align}
    \widetilde{P}(k,\tau)=e^{-D_{\text{R}} k^2 t}.
\end{align}
Plugging this in the renewal equation \eref{ren-3} we finally obtain
\begin{align}
    \widetilde{P}_{\lambda}(k,t|\theta_0)&= e^{ik\theta_0}e^{-(\lambda+D_{\text{R}} k^2) t} +\frac{\lambda}{(\lambda+D_{\text{R}} k^2) }\left(\frac{\sin k \phi}{k\phi}\right)\left[1-e^{-(\lambda+D_{\text{R}} k^2) t}\right]. \label{pk-II}
\end{align}
Let us now proceed to find the mean and variance of the ABP under this reset protocol.

\subsubsection{\textbf{Mean position}}
The mean position of the particle can be obtained from \eref{mx} and \eref{my}. Following similar steps as before, we find
\begin{align}
     \widetilde{P}_{\lambda}(k\to 1,t|\theta_0)&=\int_{-\infty}^{\infty}d\theta ~e^{i\theta}  P_{\lambda}(\theta,t|\theta_0)\nonumber\\
     &=\int_{-\infty}^{\infty}d\theta ~\cos\theta  P_{\lambda}(\theta,t|\theta_0) +i \int_{-\infty}^{\infty}d\theta ~\sin \theta  P_{\lambda}(\theta,t|\theta_0) \nonumber\\
     &=\langle \cos \theta(t) \rangle + i\langle \sin \theta (t) \rangle\nonumber\\
     &=(\cos \theta_0+i \sin \theta_0) ~e^{-(\lambda+D_{\text{R}}) t} +\frac{\lambda}{(\lambda+D_{\text{R}}) }\left(\frac{\sin \phi}{\phi}\right)\left[1-e^{-(\lambda+D_{\text{R}}) t}\right].
\end{align}
From which we obtain
\begin{align}
    \langle \cos \theta(t) \rangle&=\cos \theta_0 ~e^{-(\lambda+D_{\text{R}} t} +\frac{\lambda}{(\lambda+D_{\text{R}}) }\left(\frac{\sin \phi}{\phi}\right)\left[1-e^{-(\lambda+D_{\text{R}}) t}\right],\\
    \langle \sin  \theta (t)\rangle&=\sin \theta_0 ~e^{-(\lambda+D_{\text{R}}) t} .
\end{align}
Plugging these results back to \eref{mx} and \eref{my} we find the mean of position for this protocol
\begin{align}
    \langle x(t) \rangle_{\theta_0}&=\frac{V e^{- (D_{\text{R}}+\lambda)t} \left(  (D_{\text{R}}+\lambda) \cos \theta_0 \left(e^{t (D_{\text{R}}+\lambda)}-1\right)+ \lambda \left(\frac{\sin \phi}{\phi}\right)  \left(e^{t (D_{\text{R}}+\lambda)} (t (D_{\text{R}}+\lambda)-1)+1\right)\right)}{  (D_{\text{R}}+\lambda)^2}, \\
    \langle y(t) \rangle_{\theta_0}&=\frac{V \sin \theta_0 \left(1-e^{- (D_{\text{R}}+\lambda)t}\right)}{D_{\text{R}}+\lambda}.
\end{align}

\subsubsection{\textbf{Variance of the position}}

The second moment in the $x$-direction can be found using the same formula as in \eref{2m}. Similar to that case here we need to find the correlation $\left\langle e^{i k (\theta'+\theta'')}\right\rangle$. Let us try to evaluate that in the following
\begin{align}
   \left\langle e^{i( k'\theta'+k''\theta'')}\right\rangle &=\int_{-\infty}^{\infty} d\theta'' ~e^{ik''\theta''} \int_{-\infty}^{\infty} d\theta'~  e^{ik'\theta'}  P_{\lambda}(\theta'',t''|\theta',t') P_{\lambda}(\theta',t'|\theta_0) ~~ \text{for }~~t''>t'.
   \end{align}
Note that as the propagator is not homogeneous here, we can not use the Fourier convolution method as we did in the earlier case. However, we can still simplify our calculations in the following fashion
\begin{align}
    \left\langle e^{i( k'\theta'+k''\theta'')}\right\rangle &=\int_{-\infty}^{\infty} d\theta'~  e^{ik'\theta'}   P_{\lambda}(\theta',t'|\theta_0) \int_{-\infty}^{\infty}  d\theta''~e^{ik''\theta''}  P_{\lambda}(\theta'',t''|\theta',t') \nonumber \\
    &=\int_{-\infty}^{\infty} d\theta'~  e^{ik'\theta'}   P_{\lambda}(\theta',t'|\theta_0) \widetilde{P}_{\lambda}(k'',t''-t'|\theta').
\end{align}
Let us now plug the expression for $\widetilde{P}_{\lambda}(k'',t''-t'|\theta')$ as obtained in \eref{pk-II} to have 
\begin{align}
     &\left\langle e^{i( k'\theta'+k''\theta'')}\right\rangle \nonumber\\
     &=\int_{-\infty}^{\infty} d\theta'~  e^{ik'\theta'}   P_{\lambda}(\theta',t'|\theta_0) \left[ e^{ik''\theta'}e^{-(\lambda+D_{\text{R}} k''^2) (t''-t')} +\frac{\lambda}{(\lambda+D_{\text{R}} k''^2) }\left(\frac{\sin k'' \phi}{k''\phi}\right)\left[1-e^{-(\lambda+D_{\text{R}} k''^2) (t''-t')}\right]\right]\nonumber\\
   &=  e^{-(\lambda+D_{\text{R}} k''^2) (t''-t')} \widetilde{P}_{\lambda}(k'+k'',t'|\theta_0) + \frac{\lambda}{(\lambda+D_{\text{R}} k''^2) }\left(\frac{\sin k'' \phi}{k''\phi}\right)\left[1-e^{-(\lambda+D_{\text{R}} k''^2) (t''-t')}\right] \widetilde{P}_{\lambda}(k',t'|\theta_0).
\end{align}
Using the above relation in \eref{eit} (with setting $\{k',k''\}=\pm 1$) one can obtain the correlation function as
\begin{align}
    \langle \cos \theta(t') \cos \theta(t'') \rangle=
& \frac{1}{8 \phi^2} \Bigg[ 
-\frac{8 \lambda^2 \sin^2\phi~ e^{-(t' (D_{\text{R}}+\lambda))} \left(e^{t' (D_{\text{R}}+\lambda)}-1\right) \left(e^{(D_{\text{R}}+\lambda) (t'-t'')}-1\right)}{(D_{\text{R}}+\lambda)^2} \nonumber \\ 
& + \frac{2 \lambda \phi \sin \phi e^{-D_{\text{R}} (4 t'+t'')- \lambda (t'+t'')}}{(D_{\text{R}}+\lambda) (4 D_{\text{R}}+\lambda)} \Bigg\{
(D_{\text{R}}+\lambda) \cos\phi e^{t' (D_{\text{R}}+\lambda)} \left(e^{t' (4 D_{\text{R}}+\lambda)}-1\right) \nonumber \\ 
& -2 (4 D_{\text{R}}+\lambda) e^{3 D_{\text{R}} t'} (2 \cos\theta_0) \left(e^{t' (D_{\text{R}}+\lambda)}-e^{t'' (D_{\text{R}}+\lambda)}\right) 
\Bigg\} \nonumber \\ 
& + \frac{\phi e^{-t'' (D_{\text{R}}+\lambda)-3 D_{\text{R}} t'}}{4 D_{\text{R}}+\lambda} 
\Bigg\{ 
2 \phi (4 D_{\text{R}}+\lambda) \left(2 e^{4 D_{\text{R}} t'+\lambda t'}+2 \cos2 \theta_0\right) \nonumber \\ 
& + \lambda \sin2 \phi \left(e^{t' (4 D_{\text{R}}+\lambda)}-1\right) 
\Bigg\} 
\Bigg] .
\end{align}
Plugging this result in \eref{2m-2} we find the second moment for this protocol as
\begin{align}
\langle x^2 (t) \rangle_{\theta_0}=& \frac{V^2e^{-5 D_{\text{R}} t - 2 \lambda t}}{6 D_{\text{R}} \phi^2 (D_{\text{R}}+\lambda)^4 (4 D_{\text{R}}+\lambda)^2} \Bigg[ \nonumber \\ 
& 2 \lambda \sin(\phi) \Bigg\{
\phi (D_{\text{R}}+\lambda)^2 \cos(\phi) e^{t (D_{\text{R}}+\lambda)} 
\Big( 16 D_{\text{R}}^2 e^{3 D_{\text{R}} t} + 8 D_{\text{R}} \lambda e^{3 D_{\text{R}} t} + \lambda^2 e^{3 D_{\text{R}} t} - D_{\text{R}}^2 \nonumber \\ 
& + 3 D_{\text{R}} e^{t (4 D_{\text{R}}+\lambda)} (t (D_{\text{R}}+\lambda) (4 D_{\text{R}}+\lambda) - 5 D_{\text{R}} - 2 \lambda) - 2 D_{\text{R}} \lambda - \lambda^2 
\Big) \nonumber \\ 
& + 6 D_{\text{R}} \phi (4 D_{\text{R}}+\lambda)^2 (D_{\text{R}}+\lambda) \cos(\theta_0) e^{t (4 D_{\text{R}}+\lambda)} 
\Big( t (D_{\text{R}}+\lambda) + e^{t (D_{\text{R}}+\lambda)} (t (D_{\text{R}}+\lambda) - 2) + 2 \Big) \nonumber \\ 
& + 3 D_{\text{R}} \lambda (4 D_{\text{R}}+\lambda)^2 \sin(\phi) e^{5 D_{\text{R}} t + 2 \lambda t} 
\Big( t (D_{\text{R}}+\lambda) (t (D_{\text{R}}+\lambda) - 4) + 6 \Big) 
\Bigg\} \nonumber \\ 
& + 6 D_{\text{R}} (4 D_{\text{R}}+\lambda)^2 \Bigg\{
e^{t (4 D_{\text{R}}+\lambda)} \Big(\phi^2 (D_{\text{R}}+\lambda)^2 - \lambda^2 (t (D_{\text{R}}+\lambda) + 3) \Big) \nonumber \\ 
& + \phi^2 (D_{\text{R}}+\lambda)^2 e^{5 D_{\text{R}} t + 2 \lambda t} (t (D_{\text{R}}+\lambda) - 1) 
\Bigg\} \nonumber \\ 
& + 6 D_{\text{R}} \lambda^2 (4 D_{\text{R}}+\lambda)^2 \cos(2 \phi) e^{t (4 D_{\text{R}}+\lambda)} 
\Big( t (D_{\text{R}}+\lambda) + 3 \Big) \nonumber \\ 
& + 2 \phi^2 (4 D_{\text{R}}+\lambda) (D_{\text{R}}+\lambda)^3 \cos(2 \theta_0) e^{-t (4 D_{\text{R}}+\lambda)} 
\Bigg\{ 3 D_{\text{R}} e^{t (4 D_{\text{R}}+\lambda)} - 4 D_{\text{R}} e^{3 D_{\text{R}} t} - \lambda e^{3 D_{\text{R}} t} + D_{\text{R}}  + \lambda \Bigg\} 
\Bigg] .
\end{align}
Following a similar procedure, the second moment in the $y$-direction is found to be
\begin{align}
\langle y^2 (t) \rangle_{\theta_0}&= \frac{V^2e^{-5 D_{\text{R}} t - 2 \lambda t}}{6 D_{\text{R}} \phi (D_{\text{R}}  + \lambda)^2 (4 D_{\text{R}}  + \lambda)^2} \Bigg[ e^{t (D_{\text{R}}  + \lambda)} \Bigg\{ 
\lambda \sin(2 \phi) \Big( -16 D_{\text{R}}^2 e^{3 D_{\text{R}} t} + D_{\text{R}}^2 - \lambda^2 e^{3 D_{\text{R}} t} - 8 D_{\text{R}} \lambda e^{3 D_{\text{R}} t} \nonumber \\  
& - 3 D_{\text{R}} e^{t (4 D_{\text{R}}  + \lambda)} (t (D_{\text{R}}  + \lambda) (4 D_{\text{R}}  + \lambda) - 5 D_{\text{R}} - 2 \lambda) + 2 D_{\text{R}} \lambda + \lambda^2 \Big) \nonumber \\  
& + 2 \phi (D_{\text{R}}  + \lambda) (4 D_{\text{R}}  + \lambda) \cos(2 \theta_0) \Big( 
D_{\text{R}} (-3 e^{t (4 D_{\text{R}}  + \lambda)} + 4 e^{3 D_{\text{R}} t} - 1) + \lambda (e^{3 D_{\text{R}} t} - 1) 
\Big) \Bigg\} \nonumber \\  
& + 6 D_{\text{R}} \phi (4 D_{\text{R}}  + \lambda)^2 e^{t (4 D_{\text{R}}  + \lambda)} \Big( e^{t (D_{\text{R}}  + \lambda)} (t (D_{\text{R}}  + \lambda) - 1) + 1 \Big)  
\Bigg] .
\end{align}

\subsection{\textit{Average over initial $\theta_0$ and adding translational diffusion}} As before we consider the initial orientation of the ABP is chosen from the uniform distribution inside $\theta_0\in [0,2\pi]$. In addition we also take into account the translational diffusion in both the direction, which in turn yields the following expressions for the MSD:

\begin{figure}
    \centering
    \includegraphics[width=8cm]{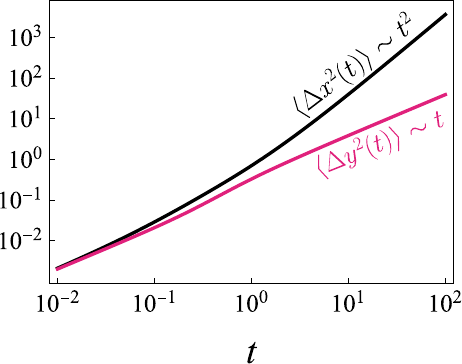}
    \caption{MSD along $x$ and $y$-direction for case-II for the parameters: $D_{\text{R}}=1/2.61,V=1.02,\lambda=5,\phi=\pi/2, D_{\text{T}}=0.1$. Note that the MSD along the $x$-direction grows ballistically with respect to time, in contrast to the $y$-MSD which has a diffusive behaviour at large enought times.}
    \label{fig:placeholder}
\end{figure}
\begin{align}
&\langle \Delta x^2(t) \rangle= \frac{V^2}{12 \phi^2 (D_{\text{R}}+\lambda)^3} \Bigg[ 
  \frac{2 \phi (D_{\text{R}}+\lambda) e^{-t (4 D_{\text{R}}+\lambda)}}{D_{\text{R}} (4 D_{\text{R}}+\lambda)^2} 
  \Bigg\{  \nonumber \\
& \quad \lambda \sin(2\phi) \Big( 
    16 D_{\text{R}}^2 e^{3 D_{\text{R}} t} - D_{\text{R}}^2 + \lambda^2 e^{3 D_{\text{R}} t} 
    + 8 D_{\text{R}} \lambda e^{3 D_{\text{R}} t}  \nonumber \\
& \qquad + 3 D_{\text{R}} e^{t (4 D_{\text{R}}+\lambda)} 
      \big( t (D_{\text{R}}+\lambda)(4 D_{\text{R}}+\lambda) - 5 D_{\text{R}} - 2\lambda \big) 
      - 2 D_{\text{R}} \lambda - \lambda^2 
  \Big)  \nonumber\\
& \quad + 2 \phi \Big( 
      32 D_{\text{R}}^3 e^{3 D_{\text{R}} t} + 4 D_{\text{R}}^3 + 9 D_{\text{R}}^2 \lambda 
      - \lambda^3 e^{3 D_{\text{R}} t} - 6 D_{\text{R}} \lambda^2 e^{3 D_{\text{R}} t}  \nonumber\\
& \qquad + 6 D_{\text{R}} \lambda^2 
      + 3 D_{\text{R}} (4 D_{\text{R}}+\lambda) e^{t (4 D_{\text{R}}+\lambda)} 
        \big( t (D_{\text{R}}+\lambda)(4 D_{\text{R}}+\lambda) - 3 D_{\text{R}} \big) 
      + \lambda^3 
  \Big) 
  \Bigg\}  \nonumber \\
& \quad + \frac{12 \lambda^2 \sin^2(\phi) e^{-t(D_{\text{R}}+\lambda)} 
      \left( -2t(D_{\text{R}}+\lambda) 
      + e^{t(D_{\text{R}}+\lambda)} \big( t(D_{\text{R}}+\lambda)(t(D_{\text{R}}+\lambda)-4)+6 \big) - 6 \right)}
    {D_{\text{R}}+\lambda}  \nonumber \\
& \quad + 24 \lambda \phi \sin(\phi) e^{-t(D_{\text{R}}+\lambda)} 
      \Big( t(D_{\text{R}}+\lambda) + e^{t(D_{\text{R}}+\lambda)} (t(D_{\text{R}}+\lambda)-2) + 2 \Big) 
\Bigg]+ 2 D_{\text{T}} t .
\label{MSD-x}
\end{align}

and

\begin{align}
& \langle \Delta y^2(t) \rangle  =  V^2 \Bigg( 
\frac{1}{12 D_{\text{R}} (D_{\text{R}}+\lambda)(4 D_{\text{R}}+\lambda)\phi}
\Bigg[
\Bigg( 
-\frac{2 e^{-(4 D_{\text{R}}+\lambda) t} (D_{\text{R}}+\lambda)}{4 D_{\text{R}}+\lambda} 
+ \frac{2 e^{-(D_{\text{R}}+\lambda) t} (4 D_{\text{R}}+\lambda)}{D_{\text{R}}+\lambda} \nonumber \\
& \qquad 
+ 6 D_{\text{R}} \left( -\frac{1}{D_{\text{R}}+\lambda} - \frac{1}{4 D_{\text{R}}+\lambda} + t \right) 
\Bigg) \times \big( 2(4 D_{\text{R}}+\lambda)\phi - \lambda \sin(2\phi) \big)
\Bigg] \Bigg)  + 2 D_{\text{T}} t.
 \label{MSD-y}
\end{align}
The \eref{MSD-x} and \eref{MSD-y} are plotted in Fig. (4) of the main text. Note that in the limit $t\to \infty (\gg \lambda+D_{\text{R}})$ the above expressions as the following asymptotic form
\begin{align}
   \langle \Delta x^2(t\to\infty) \rangle \sim  \left[\frac{V^2\lambda^2  \sin ^2(\phi )}{\phi ^2 (D_\text{R}+\lambda)^2}\right] t^2, ~~~ \langle \Delta y^2(t\to\infty) \rangle \sim \left[ \frac{2 D_\text{T} (D_\text{R}+\lambda)-\frac{\lambda V^2 \sin (\phi ) \cos (\phi )}{4 D_\text{R} \phi +\lambda \phi }+V^2}{D_\text{R}+\lambda}\right] t.
\end{align}

\newpage
\bibliography{fpusr}